\documentclass[
 amsmath,amssymb,
 aps, twocolumn,
 pra] {revtex4-1}

\usepackage     {graphicx}      
\usepackage     {color}         
\usepackage		{upgreek}       
\usepackage     {hyperref}      
\usepackage     {braket}        
\usepackage		{amsmath}		
\usepackage		{siunitx}		
\usepackage     {float}         
\usepackage     {gensymb}       
\usepackage	[capitalize]	{cleveref}		

\Crefname{figure}{fig.}{Figures}
\Crefname{fig_a}{fig.}{Figures}
\Crefname{fig_b}{fig.}{Figures}
\Crefname{fig_c}{fig.}{Figures}

\definecolor{green}{rgb}{0.45, 0.45, 0}
\definecolor{darkred}{rgb}{0.5, 0, 0}
\definecolor{gray}{gray}{0.4}
\hypersetup{
    colorlinks,
    linkcolor=gray,
    citecolor=darkred,
    urlcolor=green,
    pdftitle={Unravelling the Structures in the van der Waals Interactions of Alkali Rydberg Atoms},
    pdfauthor={K. Wadenpfuhl and C. S. Adams},
}

\begin{document}


\title{Unravelling the Structures in the van der Waals Interactions of Alkali Rydberg Atoms}

\author{Karen Wadenpfuhl}
\email{karen.wadenpfuhl@durham.ac.uk}
\author{C. Stuart Adams}
\email{c.s.adams@durham.ac.uk}
\affiliation{Joint Quantum Centre (JQC) Durham-Newcastle, Department of Physics, Durham University, DH1 3LE, United Kingdom}

\date{\today}

\begin{abstract}
Rydberg atoms are used in a wide range of applications due to their peculiar properties like strong dipolar and van der Waals interactions. The choice of Rydberg state has a huge impact on the strength and angular dependence of the interactions, and so a detailed understanding of the underlying processes and resulting properties of the interactions is therefore key to select the most suitable states for experiments. We study the van der Waals interactions in alkali atoms in detail and highlight the structures which allow an understanding and exploitation of the various interaction properties. A particular theme is the identification of F\"orster resonances with $n_1 \neq n_2$, which offer interaction potentials with a wide range of properties that make them particularly interesting for experimental applications. A second theme is a focus on the underlying structures that shape the angular dependency and sign of the interactions. This understanding -- instead of brute-force calculations -- allows for a much simpler and more systematic search for suitable pair states. These insights can be used for the selection of tailored interaction potentials subject to experimental constraints and requirements. We use rubidium as an example species in this work and also provide data for cesium and pair states that are coupled via two- or three-photon transitions, i.e. up to F states, in the appendix.
\end{abstract}

\maketitle

The versatility and exaggerated properties of Rydberg atoms have seen them become a favourite toy of many atomic physicists since their extreme properties greatly enhanced the toolbox of atomic physics \cite{White1934, Gallagher1994}. Nowadays, Rydberg atoms are used in a wide range of applications, such as electromagnetic field sensing \cite{Sedlacek2013, Chen2022, Fan2015, Schlossberger2024}, the probing of fundamental physical constants \cite{Jentschura2008, Scheidegger2024}, and quantum computing \cite{Wu2021, Evered2023, Anand2024, Saffman2010, Saffman2016} or simulation \cite{Bernien2017, Scholl2021, Chen2023, Geier2024, Browaeys2020, Weimer2010}. The latter in particular harness the strong dipole or van der Waals interactions between Rydberg atoms in close spatial proximity, such as Rydberg CNOT gates in neutral-atom based quantum computing \cite{Jaksch2000, Isenhower2010}. In addition, the mutual interactions in Rydberg systems have been used to study glassy dynamics arising in disordered spin models \cite{Signoles2021, Schultzen2022} or time-reversal dynamics \cite{Geier2024}. Furthermore, Rydberg interactions lead to optical nonlinearities \cite{Pritchard2010, Peyronel2012, Busche2017, Tebben2021, Srakaew2023, Firstenberg2016}, self-organisation \cite{Ding2020} and other complex dynamical behaviour \cite{Wadenpfuhl2023, Wu2024, Helmrich2020, Klocke2021}. 

Irrespective of the nature of the experiment, it is key to understand Rydberg interactions in detail in order to select interaction potentials tailored to experimental needs. In this work, we provide a detailed investigation of the second-order van der Waals (vdW) interactions, which is often relied upon in Rydberg experiments. Disentangling the different contributions to the interaction strength, namely radial and angular coupling, as well as the energy defect, leads to the formulation of angular channels. The associated structure of these angular channels provides an intuitive and straightforward explanation for the angular dependency of the resulting vdW interactions. This picture facilitates  the prediction and selection of interaction potentials required for any given experiment. Additionally, we show how Rydberg pair states with different principal quantum numbers $n_1 \neq n_2$ can have particularly pronounced F\"orster resonances, which makes them strong candidates for future experiments. Our approach explains why F\"orster resonances usually lead to strong angular dependencies in the vdW interaction, and we show how one can easily identify pair states with strong or vanishing angular dependencies.

Previous systematic investigations of the vdW interactions have either been restricted to pair states with $n_1 = n_2$, e.g. \cite{Walker2005, Walker2008, Reinhard2007}, or relied on brute-force calculations, such as \cite{Weber2017, Sibalic2017}. Focusing on Rydberg states with $n_1 = n_2$ is a limitation now that ever more experiments use multiple Rydberg excitation lasers. Approaches based on brute-force calculations, on the other hand, do not provide detailed insights into the fundamental reasons for the angular dependencies and F\"orster resonance structures. We aim to alleviate these shortcomings with this work.

In order to demonstrate our approach, we have chosen $\ket{n_1 P_{1/2}, n_2 P_{1/2}}$ pair states even though they are not directly addressable via two-photon excitation schemes as used in many Rydberg experiments. The motivation for choosing these pair states as our example is that they couple to only four angular channels. They still display a rich internal structure of F\"orster resonances and angular dependecies, and hence allow a simple but clear demonstration of our approach.

\section{Dipole-Dipole Interactions}
\label{sec:dipoledipoleInteractions}

Atoms possess temporary dipole moments while transitioning between different electronic states \cite{Scully1997}, which means that Rydberg atoms in spatial proximity can interact electromagnetically even though appearing electrically neutral at large distances. The resulting interaction can be calculated via a multipole expansion \cite{Saffman2010, Browaeys2016, Sibalic2017, Weber2017}, which is well-justified for non-overlapping charge distributions. The leading term of the multipole expansion is given by the interaction between two dipoles $\hat{V}_{dd}(\textbf{R})$. 

\begin{figure}
\centering
\includegraphics[width=\linewidth]{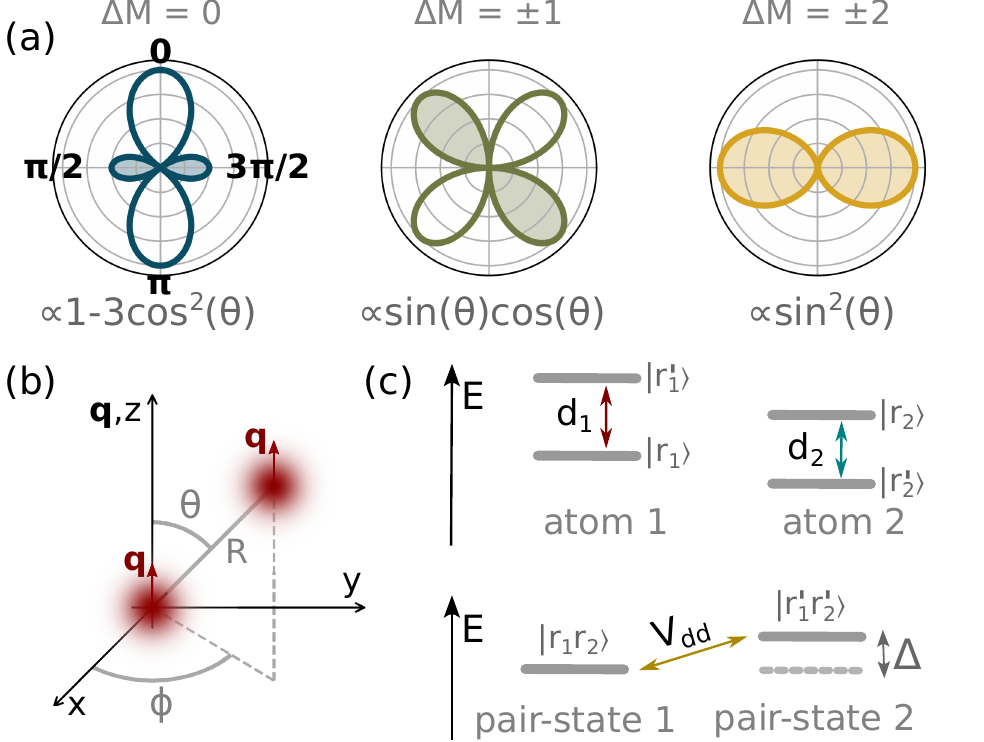}
\caption{\textbf{Dipole-dipole interactions.}
(a) shows the angular dependency of $\hat{V}_{dd}$ for processes with $\Delta M = 0, \pm 1, \pm 2$. Shaded areas have positive sign. (b) Definition of the relative position $(R, \theta, \phi)$ in a spherical coordinate system with the quantisation axis $\textbf{q}\ ||\ z$. (c) Single-atom (top) and pair-state (bottom) picture of the dipole-coupled system. $\Delta$ indicates the energy difference (defect) between the pair states.}
\label{fig:dipoleSpherical}
\end{figure}

Assuming, for simplicity, that the $z$-axis is parallel to the quantisation axis \textbf{q} one finds the well-known expression of the dipole operator in spherical coordinates \cite{Ravets2015}
\begin{equation}
\label{eqn:dipoleInteractionSpherical}
\begin{split}
& \hat{V}_\mathrm{dd}(\textbf{R}) = \frac{1}{4\pi\epsilon_0 R^3} \times \\
& \hspace{1em} \times \left[
\begin{array}{c}
\left(1-3\cos^2(\theta)\right) \left[\hat{d}_i^0 \hat{d}_j^0 + \frac{1}{2}(\hat{d}_i^+\hat{d}_j^- +\hat{d}_i^-\hat{d}_j^+) \right] \\[.8em]
-\frac{3}{\sqrt{2}}\sin(\theta ) \cos(\theta ) \left[\begin{matrix}
+e^{+i\phi}(\hat{d}_i^0\hat{d}_j^- +\hat{d}_i^-\hat{d}_j^0) \\[.25em] - e^{-i\phi}(\hat{d}_i^0\hat{d}_j^+ +\hat{d}_i^+\hat{d}_j^0) \end{matrix}\right] \\[1.2em]
-\frac{3}{2}\sin^2(\theta ) \left[e^{+2i\phi}\hat{d}_i^-\hat{d}_j^- + e^{-2i\phi}\hat{d}_i^+\hat{d}_j^+\right]
\end{array}
\right]
\end{split}
\end{equation}
with the dipole operators defined as $\hat{d}^0 = \hat{d}^z$ and $\hat{d}^\pm~=~\mp~\frac{1}{\sqrt{2}}\left(\hat{d}^x~\pm~i\hat{d}^y\right)$ for coupling to linearly and circularly polarised light respectively. The terms in the upper row result in no change of the total magnetic quantum number $M = m_1 + m_2$, while the middle and lower row lead to changes of $\Delta M = \pm 1$ and $\Delta M = \pm 2$. The change in total magnetic quantum number $\Delta M$ therefore determines the angular dependency of the processes in $V_{dd}$, as show in Figure \ref{fig:dipoleSpherical} (a).

This dipole-operator quantifies the leading term of the interaction between two Rydberg atoms with orientation $(R, \theta, \phi)$, as defined in Figure \ref{fig:dipoleSpherical} (b), while transitioning from $\ket{r_1} \to \ket{r_1^\prime}$ and $\ket{r_2} \to \ket{r_2^\prime}$, as shown in Figure \ref{fig:dipoleSpherical} (c). Effectively, the dipole-dipole interaction leads to a coupling $V(\textbf{R}) = \bra{r_1, r_2} \hat{V}_{dd}(\textbf{R}) \ket{r_1^\prime, r_2^\prime}$ of the pair states $\ket{r_1, r_2}$ and $\ket{r_1^\prime, r_2^\prime}$, which have an energy difference $\Delta = E(r_1, r_2) - E(r_1^\prime, r_2^\prime)$.

For interatomic distances below the vdW radius $R_{vdW} = \sqrt[3]{|V(\textbf{R})/ \Delta|}$, the interactions scale as $\propto V(\textbf{R}) = C_3(\textbf{R}) / R^3$ and grow with $(n^\star)^4$ \cite{Saffman2010}, where $n^*$ denotes the effective principal quantum number. In the vdW limit with $r \gg R_{vdW}$, on the other hand, the interaction-induced level shift scales as $\propto V(\textbf{R})^2 = C_6(\textbf{R}) / R^6$ and grows with the effective principal quantum number $(n^\star)^{11}$ \cite{Saffman2010, Walker2008}. In this work, we consider the vdW limit where $V(\textbf{R}) \ll \Delta$ such that second-order perturbation theory is applicable.

\section{Van der Waals Interactions}
\label{sec:vdwaalsinteraction}

\begin{figure*}
\centering
\includegraphics[width=\linewidth]{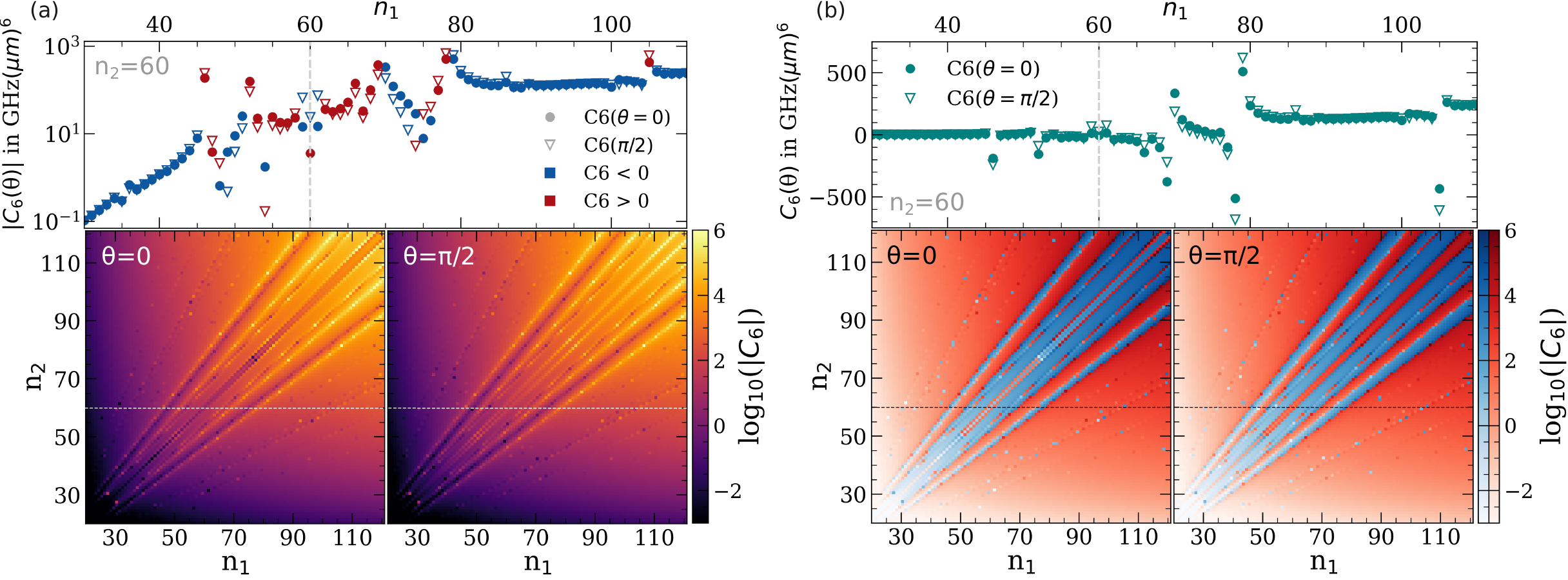}
\caption{\textbf{Structures in $\mathbf{C_6(\theta)}$ values of $\mathbf{\ket{n_1P_{1/2}, n_2P_{1/2}}}$ in rubidium.} (a) shows the absolute value of $C_6(\theta)$ and (b) the sign of $C_6(\theta)$ for $\ket{n_1P_{1/2}, n_2P_{1/2}}$ pair states in rubidium (red: $C_6 > 0$, blue: $C_6 < 0$). The 2D maps on the LHS in (a) and (b) show $C_6(\theta)$ at $\theta=0$ while the maps on the RHS show the respective same but for $\theta = \pi/2$. The top row plot shows a cross-section through the corresponding 2D maps at the dashed lines where $n_2 = 60$. Several lines of strong resonances (bright lines in a) can be seen, which are accompanied by a sign (colour) change in (b). The difference in $C_6$ for $\theta = 0, \pi/2$ is strongly pronounced at e.g. $n_1 \approx n_2$, which is indicated by the open triangles/full circles having different values. Order-of-magnitude and sign maps for other $\ket{n_1 L^\prime_{j_1}, n_2 L^{\prime\prime}_{j_2}}$ pair states for Rb and Cs can be found in Appendices \ref{app:otherStateMapsRb} and \ref{app:otherStateMapsCs} respectively.}
\label{fig:structuresInC6}
\end{figure*}

Second-order processes take place via an intermediate pair state $\ket{r_1^\prime, r_2^\prime}$ \cite{Chew2022}, but for arbitrary initial states $\ket{r_1, r_2}$ one has to take many pairs of intermediate states into account which all contribute to the resulting interaction. The vdW Hamiltonian therefore sums over all dipole-coupled intermediate pair states $\ket{r_1^\prime, r_2^\prime}$
\begin{equation}
\label{eqn:secondOrderHamiltonian}
\hat{H}(\textbf{R}) = \displaystyle\sum_{ \{ \ket{r_1^\prime, r_2^\prime} \} } \frac{\hat{V}_{dd}(\textbf{R}) \ket{r_1^\prime, r_2^\prime}\bra{r_1^\prime, r_2^\prime} \hat{V}_{dd}(\textbf{R})}{\Delta(r_1, r_2;\ r_1^\prime, r_2^\prime)}.
\end{equation}
With this Hamiltonian, the resulting interaction strength of a given pair state $\ket{r_1, r_2}$ can be calculated via 
\begin{equation}
\label{eqn:C6fromHamiltonian}
-C_6(\textbf{R};\ r_1, r_2) / R^6 = \bra{r_1, r_2}\hat{H}(\textbf{R}) \ket{r_1, r_2}.
\end{equation}
As shown in Figure \ref{fig:structuresInC6}, one finds structures arising in $C_6$ for variable $(n_1,\ n_2)$, as well as a pronounced angular dependency for certain pair states and changes in sign of $C_6$. These anisotropies and sign changes occur in the vicinity of the lines of strong resonance in Figure \ref{fig:structuresInC6} (a).

The origin of these structures and angular dependencies can be found in the Hamiltonian \eqref{eqn:secondOrderHamiltonian}. Of all possible intermediate pair states in the sum, only a few will contribute significantly to the vdW interactions. Finding these relevant intermediate states and predicting the resulting properties of the interaction for any pair state $\ket{r_1, r_2}$ is strongly simplified by isolating the angular channels $(l_1, j_1; l_2, j_2) \leftrightarrow (l_1^\prime, j_1^\prime; l_2^\prime, j_2^\prime)$ allowed in the process, and considering their respective energy defect structures. For this, we first need to have a closer look at the numerator in equation \eqref{eqn:secondOrderHamiltonian}, as this leads to the angular channel formulation of $\hat{H}(\textbf{R})$, and then at the energy defect $\Delta$ in the denominator.

\subsection{Angular Channels}
\label{subsec:angularChannels}

The second-order process consists of two consecutive dipole transitions $\hat{V}_{dd}(\textbf{R})$ with their respective dipole matrix elements in single-atom basis
\[ \begin{split}d^\alpha & = \braket{n^\prime, l^\prime, j^\prime, m_j^\prime|\hat{d}^\alpha|n, l, j, m_j} \\ & = \mathcal{R}(n, l, j;\ n^\prime, l^\prime, j^\prime) \mathcal{D}^\alpha(l, j, m_j;\ l^\prime, j^\prime, m_j^\prime)
\end{split} \]
separating into radial and angular components $\mathcal{R}$ and  $\mathcal{D}^\alpha$ respectively \cite{Sibalic2017, Walker2008, ParisMandoki2016}. Explicit expressions for each are given in Appendix \ref{app:radialAngularExpressions}.

The angular coupling $\mathcal{D}^\alpha$ depends on all quantum numbers except for the principal quantum number $n$, while the radial coupling strength depends only on $n$, $l$ and $j$. The  energy defect $\Delta$ is also independent of $m_j$, unless electric or magnetic fields are applied. This allows us to separate the sum over all intermediate state principal quantum numbers $(n_1^\prime,\ n_2^\prime)$ from the sums over the angular quantum numbers $(l^\prime_i,\ j^\prime_i)$ and re-cast the Hamiltonian from equation \eqref{eqn:secondOrderHamiltonian} into the form
\begin{widetext}
\begin{equation}
\label{eqn:angularChannels}
\hat{H}(R, \theta, \phi)\ =\ \frac{1}{R^6} \displaystyle\sum_{ \{ (l^\prime_i j^\prime_i) \} } \left( \displaystyle\sum_{ \{ (n^\prime_i) \} } \frac{\mathcal{R}_1 \mathcal{R}_2 \mathcal{R}_1^\prime \mathcal{R}_2^\prime}{\Delta(n^\prime_i, l^\prime_i, j^\prime_i)} \right) \hat{\mathcal{D}}(l^\prime_i, j^\prime_i, m_j^\prime;\ \theta, \phi) 
=\ \frac{-1}{R^6}\displaystyle\sum_{ \{ (l^\prime_i j^\prime_i) \} } C_6^{(l^\prime_i j^\prime_i)}\  \hat{\mathcal{D}}(l^\prime_i, j^\prime_i, m_i^\prime;\ \theta, \phi).
\end{equation}
\end{widetext}
All $m_j$-state dependence and angular dependencies have been absorbed in $\hat{\mathcal{D}}(l^\prime_i, j^\prime_i, m_j^\prime;\ \theta, \phi)$, as detailed in Appendix \ref{app:secondOrderAtThetaZero}, and we set $(l^\prime_i j^\prime_i) = (l^\prime_1 j^\prime_1, l^\prime_2 j^\prime_2)$ as shorthand notation.

\begin{figure}
\centering
\includegraphics[width=\linewidth]{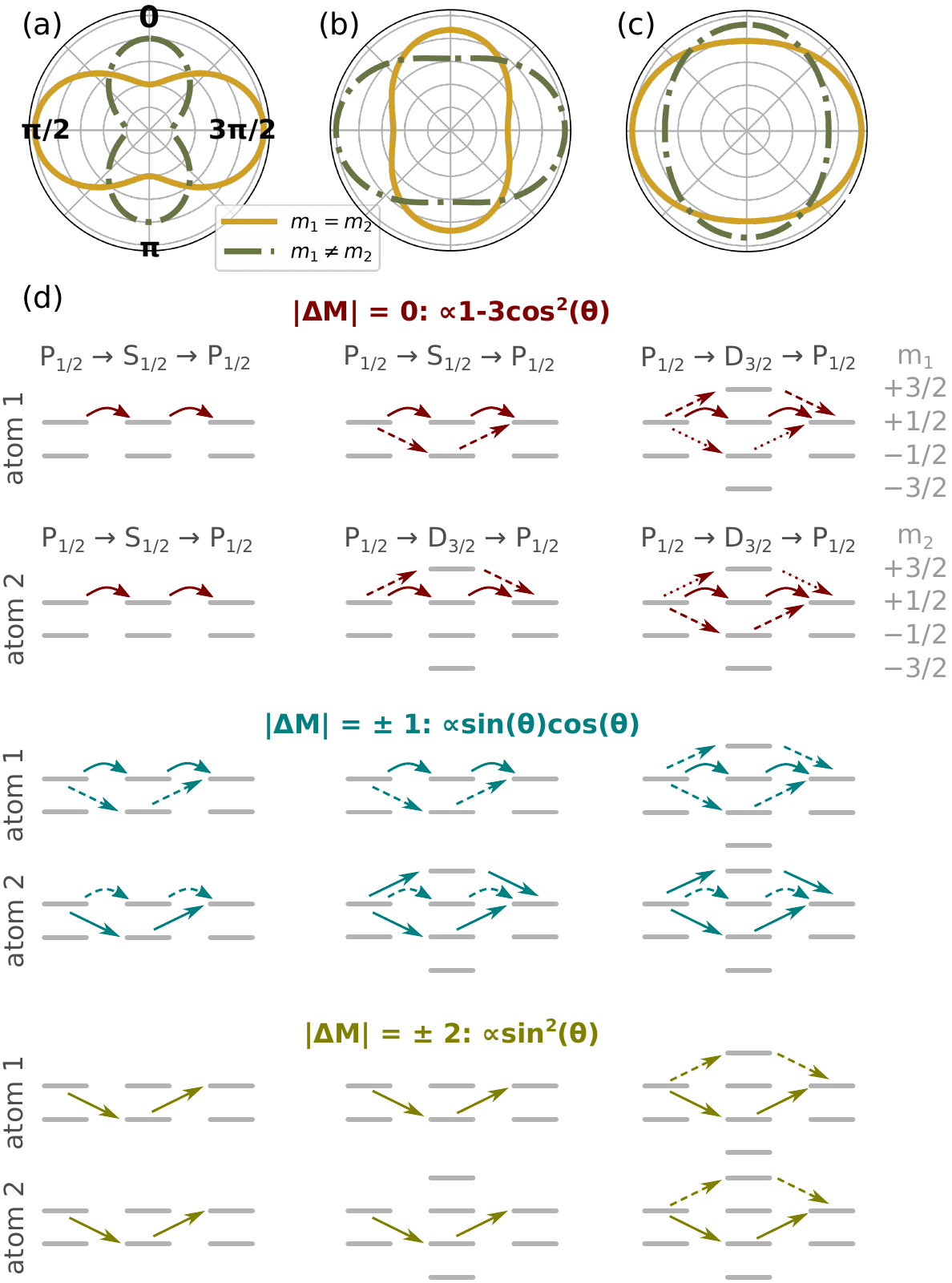}
\caption{\textbf{Angular momentum channels for $\mathbf{\ket{P_{1/2}, P_{1/2}}}$ states.} For initial and final state being $\ket{P_{1/2}, P_{1/2}}$, one has four angular momentum channels with intermediate states of the type (a) $\ket{S_{1/2}, S_{1/2}}$, (b) $\ket{S_{1/2}, D_{3/2}}$ and its permutation, and (c) $\ket{D_{3/2}, D_{3/2}}$. The resulting angular dependency of each channel is shown along $\theta$ for $m_1 = m_2$ (solid) and $m_1 \neq m_2$ (dashdot) with $\tilde{m}_i = m_i$. (d) Level schematics showing the possible $m_i \to m_i^\prime \to \tilde{m}_i$ pathways fulfilling the condition $m_i = \tilde{m}_i$, i.e. the pathways contributing to the solid lines in the respective polar plots of the angular channels. One can see that different components of $\hat{V}_{dd}$ in equation \eqref{eqn:dipoleInteractionSpherical} contribute differently to the different channels. The different total angular momentum changes $\Delta M$ of the two-atom system are shown separately, and different possible paths have different linestyles. The different $m_j$ paths available for every channel for a given set $(m_i, \tilde{m}_i)$ carry different angular dependencies and therefore result in different overall channel angular dependencies.}
\label{fig:angularMomentumChannels}
\end{figure}

Each angular momentum channel
\begin{align} \notag
\ket{l_1, j_1; l_2, j_2} \to \ket{l_1^\prime, j_1^\prime; l_2^\prime, j_2^\prime} \to \ket{l_1, j_1; l_2, j_2}
\end{align}
has a corresponding channel strength $C_6^{(l^\prime_i j^\prime_i)}$ which is, per definition, independent of the interatomic orientation $(R,\ \theta,\ \phi)$. The channel's unique angular dependency $\hat{\mathcal{D}}(l_i^\prime, j_i^\prime, m_i^\prime; \theta, \phi)$ is governed by the accessible $m_j$ substate path $(m_1, m_2) \to (m_1^\prime, m_2^\prime) \to (\tilde{m}_1, \tilde{m}_2)$, weighted by their relative strengths as in equation \eqref{eqn:dipoleInteractionSpherical}. This is shown in Figure \ref{fig:angularMomentumChannels} for a $\ket{P_{1/2}, P_{1/2}}$ state, which effectively has three different angular momentum channels, up to a permutation of the $\ket{S_{1/2}, D_{3/2}}$ channel. For different intermediate states, we have different $m_j$ pathways accessible when starting in a given initial $(m_1, m_2)$ pair state and ending in $(\tilde{m}_1, \tilde{m}_2)$. This leads to different angular dependencies and coupling strengths for the various channels.

The overall angular dependency of $C_6$ depends on the relative weighting of the channels to another. If one of the channels had a much larger channel strength $C_6^{(l^\prime_i j^\prime_i)}$ than the others, then the resulting overall $C_6$ interaction were dominated by this particular channel. For this reason, we will now have a more detailed look at the channel strengths $C_6^{(l^\prime_i j^\prime_i)}$ to be able to predict where such resonances of specific angular channels occur.

As we have seen in equation \eqref{eqn:angularChannels}, the channel strength is given by a sum over the radial couplings of all possible intermediate states divided by the energy defect of the respective intermediate pair state. The radial coupling strength is governed by the overlap integral of the radial wavefunctions of states $\ket{r_i}$ and $\ket{r_i^\prime}$, which depends only on the $n$ and $l$ quantum numbers. Generally, the radial coupling strength $\mathcal{R}$ grows with increasing principal quantum number $n$. This can be seen in larger $C_6$ values for larger principal quantum numbers, see e.g. Figure \ref{fig:structuresInC6}.

\subsection{Energy Defect Structure}
\label{subsec:energydefect}

Figure \ref{fig:structuresInC6} also clearly shows lines of strong resonances and changes in the sign of $C_6$. These are caused by the energy defect $\Delta(n^\prime_i, l^\prime_i, j^\prime_i)$. Small energy defects for a given intermediate state $\ket{r_1^\prime, r_2^\prime}$ lead to strong contributions to the channel strength $C_6^{(l^\prime_i j^\prime_i)}$. In the extreme case of a near F\"orster resonant pair state, i.e. for a very small energy defect, the channel strength would be dominated by the contribution of this near F\"orster-resonant intermediate pair state alone.

\begin{figure*}
\centering
\includegraphics[width=\linewidth]{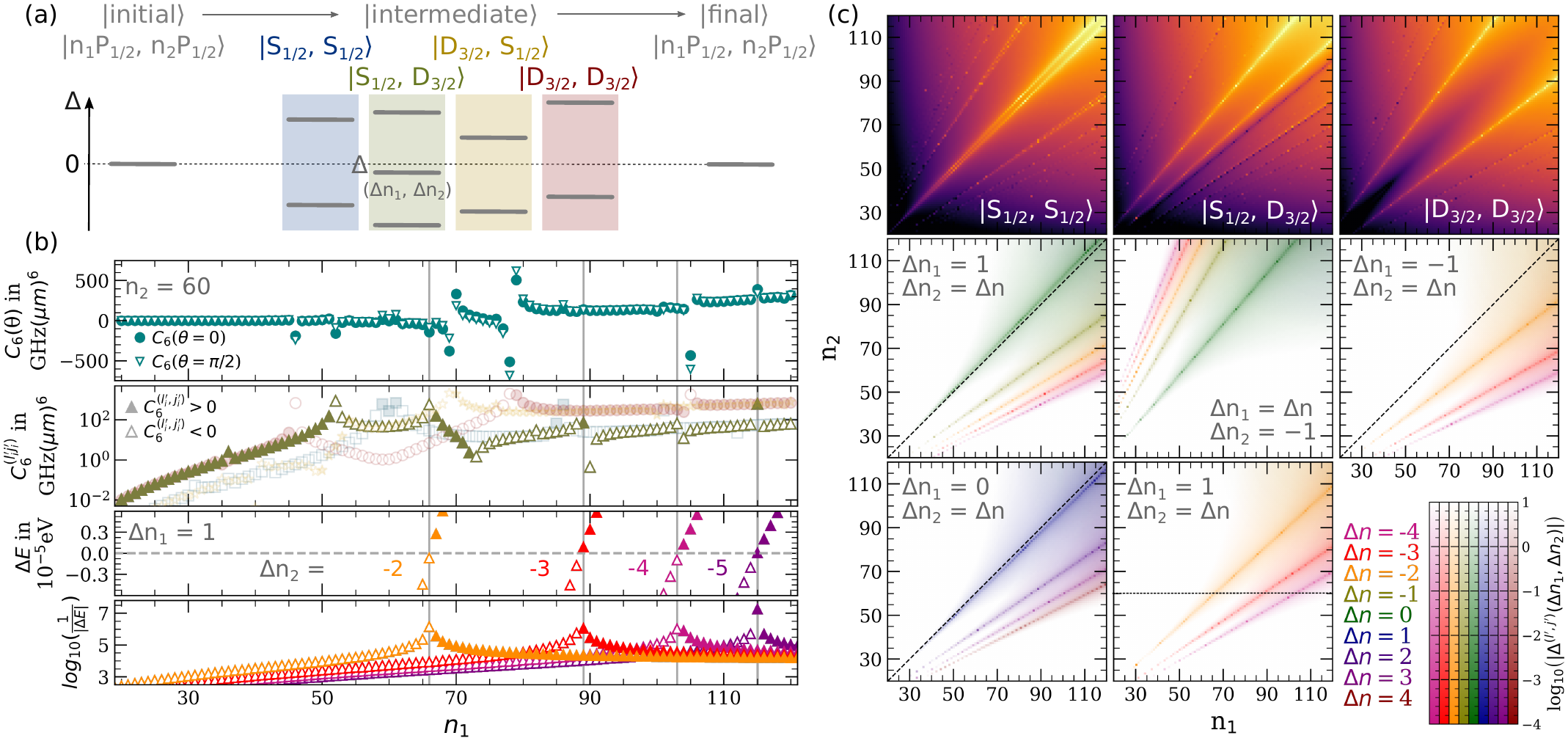}
\caption{\textbf{Relevant energy defect structures for angular momentum channels of $\mathbf{\ket{n_1 P_{1/2}, n_2 P_{1/2}}}$ states in rubidium.} For the case of $\ket{n_1 P_{1/2}, n_2 P_{1/2}}$ states, there are four possible angular channels: $\ket{S_{1/2}, S_{1/2}}$, $\ket{S_{1/2}, D_{3/2}}$, $\ket{D_{3/2}, S_{1/2}}$, $\ket{D_{3/2}, D_{3/2}}$, as shown in (a). For each angular channel, the principal quantum numbers $n_1$, $n_2$ can change up or down as the initial state couples to intermediate pair states of the form $\ket{(n_1 + \Delta n_1) L^\prime_{j_1^\prime}, (n_2 + \Delta n_2) L^{\prime\prime}_{j_2^\prime}}$. In (b) we show contributions of each angular channel to $C_6$. The data is color coded as in (a) with only the $\ket{S_{1/2}, D_{3/2}}$ channel in bold. The bottom two panels of (b) show the energy defect for $\Delta n_1 = 1$ and varying $\Delta n_2$, causing multiple lines of resonance. Note that we see that even $\Delta n_2 = -5$ contributes to the $C_6$ value. Panel (c) shows a breakdown of the contribution from different $\Delta n_i$ values for each angular channel. For the $\ket{S_{1/2}, S_{1/2}}$ channel we consider $\Delta n_1 = 0,\ 1$, see the first column in (c) and Appendix \ref{app:FoersterResonanceLines} for more details. In the plot we allow $\Delta n_2$ to take values up to $\pm 4$, as indicated by the colour. For the $\ket{S_{1/2}, D_{3/2}}$ channel, it matters whether the first or second atom has a non-zero $\Delta n$. The cases with $\Delta n_2 = -1$ and $\Delta n_1 = 1$ lead to F\"orster resonances, with the latter also shown in detail in panel (b). The $\ket{D_{3/2}, D_{3/2}}$ channel is analogous to the $\ket{S_{1/2}, S_{1/2}}$ channel. For the symmetric channels $\ket{S_{1/2}, S_{1/2}}$ and $\ket{D_{3/2}, D_{3/2}}$, swapping atoms 1 and 2 means that the figures  are simply reflected about the dashed line. Lists containing the $\Delta n_i$ combinations that cause F\"orster resonances for different channels of initial pair states up to $\ket{n_1 F_J, n_2 F_{J^\prime}}$ are listed in Appendix \ref{app:FoersterResonanceLines}.}
\label{fig:energyDefect}
\end{figure*}
Figure \ref{fig:energyDefect} shows the energy defect structure for $\ket{n_1 P_{1/2}, n_2 P_{1/2}}$ states in rubidium. For every channel, one can see that the lines of large channel strengths $|C_6^{(l^\prime_i j^\prime_i)}|$ are caused by small energy defects with intermediate pair states $\ket{(n_1 + \Delta n_1) L^\prime_{j_1^\prime}, (n_2 + \Delta n_2) L^{\prime\prime}_{j_2^\prime}}$. For different principal quantum numbers $(n_1, n_2)$ it is different changes in principal quantum number $\Delta n_i$ which minimise the energy defect, as shown in detail in Figure \ref{fig:energyDefect} (b) and (c). These lines of strong resonance lie off the $n_1 = n_2$ axis that is commonly used in experiments, and small energy defects can cause very strong vdW interactions even for very different principal quantum numbers. Additionally, the crossing of a F\"orster resonance for a given $\Delta n_1,\ \Delta n_2$ pair can also be seen in the associated sign change of the channel strength $C^{(l_i, j_i)}$.

\subsection{Combining Angular Structure and Energy Defect}
\label{subsec:combiningangularchannelsandenergydefect}

We now have all the necessary tools to understand the structure and angular dependency of the vdW interaction $C_6(\theta, \phi)$. The angular dependency of the resulting $C_6$ interaction depends on the relative strength of each angular momentum channel and its respective angular dependency. The strength of a channel's contribution is determined by the radial coupling, which increases with increasing principal quantum number $n$, and the energy defect structure. The strong structuring of $C_6$ and its changes in sign are caused by F\"orster resonances of the energy defect $\Delta$, i.e. the denominator in equation \eqref{eqn:angularChannels}. It is important to point out that the absolute value of $C_6$ depends on the azimuthal angle $\theta$ via the channel's angular dependency $\hat{\mathcal{D}}(l_i^\prime, j_i^\prime, m_i^\prime)$. The equatorial angle $\phi$ merely contributes a $\Delta M$-dependent complex phase factor that can be neglected when considering interaction strengths \cite{Reinhard2007}.

\begin{figure*}
\includegraphics[width=\linewidth]{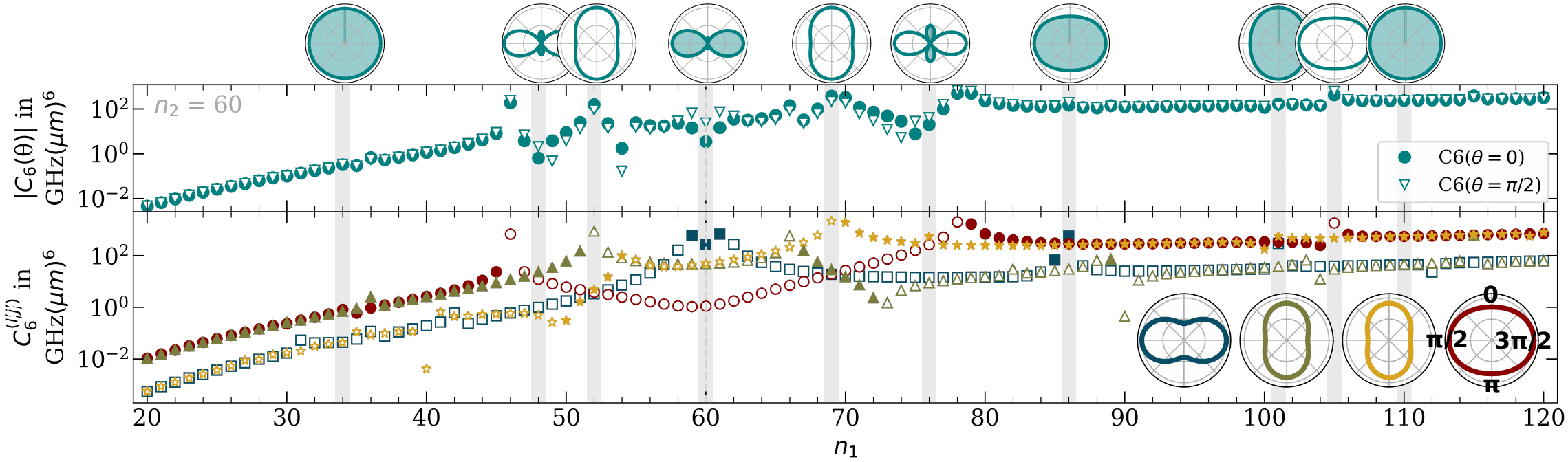}
\caption{\textbf{Channel structure and resulting angular dependency for $\mathbf{\ket{n_1 P_{1/2}, 60 P_{1/2}}}$ states in rubidium.} The upper panel shows $C_6(\theta)$ in (GHz~$\upmu$m$^6$) for $m_1 = m_2 = \pm j$ and $\tilde{m}_i = m_i$ at two different angles, $\theta = 0, \pi /2$ (solid and open symbols, respectively). The lower panel shows the corresponding structure of the four angular channels with positive (negative) channel values indicated by closed (open) symbols. The separate F\"orster resonances of the channels from Figure \ref{fig:energyDefect} are clearly visible as local peaks. The four insets in the lower plot show the angular dependencies of each channel for $m_1 = m_2 = \pm j$ and $\tilde{m}_i = m_i$ (blue: $\ket{S_{1/2}, S_{1/2}}$, green: $\ket{S_{1/2}, D_{3/2}}$, yellow: $\ket{D_{3/2}, S_{1/2}}$, red: $\ket{D_{3/2}, D_{3/2}}$). It is interesting to note that the interaction strengths for e.g. $n_1 = 52,\ n_2 = 60$ are larger than for $n_1 = n_2 = 60$ and the angular anisotropy is significantly less pronounced, as can also be seen in the insets above the main plot showing $C_6(\theta)$. Due to the resulting interaction properties, these F\"orster resonances with $n_1 \neq n_2$ are very interesting for potential experimental applications. States like e.g. $n_1 = 48,\ n_2 = 60$ show F\"orster zeroes \cite{Walker2005}, which can be exploited for orders-of-magnitude differences in interaction strength.}
\label{fig:combinedKnowledge}
\end{figure*}
Putting all of this together, we can now explain the angular dependencies of the different $\ket{n_1 P_{1/2}, 60 P_{1/2}}$ states of rubidium in Figure \ref{fig:combinedKnowledge}. The upper panel shows the absolute values of $C_6$ at interatomic orientations $\theta = 0, \pi/2$ relative to the quantisation axis. The lower panel shows the corresponding channel strengths $C_6^{(l^\prime_i j^\prime_i)}$ with their various F\"orster resonances visible as local peaks in the channel strengths. The insets on the bottom right show the channel's angular dependencies for $m_i = \tilde{m}_i = \pm 1/2$, and the insets on top show the corresponding $C_6(\theta )$ values. $n_1 = 52$ is an example where the $\ket{S_{1/2}, D_{3/2}}$ channel dominates the others by orders of magnitude and therefore detemines the angular dependency. F\"orster resonances of an angular channel are therefore usually associated with an associated angular dependency of $C_6(\theta)$, and sign flips of $C_6$ occur in the vicinity of channel resonances. A counterexample is set by e.g. $n_1 = 110$, where the angular dependencies of the different channels cancel which leads to effectively isotropic interactions. Pair states like $\ket{76P_{1/2}, 60P_{1/2}}$, on the other hand, show that the angular dependency of $C_6(\theta)$ is the sum of the weighted angular dependencies of the different channels. Here, two channels of different sign cancel each other at a F\"orster zero \cite{Walker2005} angle where the interaction strength vanishes as the sign of $C_6$ changes along $\theta$.

\section{Conclusion}
\label{sec:conclusion}

In this paper we have highlighted the importance of disentangling the different contributions to the van der Waals interaction potential, i.e. the angular coupling from the radial part and the energy defect. Separating radial and angular parts allows us to phrase the interactions in terms of angular channels, which each bear a characteristic angular dependency that follows from the accessible $m_j$ paths. The strength of an angular channel is determined by the radial coupling and energy defect structure of the intermediate pair states. F\"orster resonances in the energy defect lead to strong contributions to the channel strength $C_6^{(l_i, j_i)}$. When adding the different angular channels, one arrives at the overall interaction potential $C_6(\theta)$. In the vicinity of F\"orster resonances, the properties of $C_6$ show structures such as exaggerated interaction strengths and strong angular dependencies as well as changes in sign of $C_6$.

Our angular channel approach can therefore be used to identify pair states with strong or vanishing angular dependency to match experimental requirements. As the calculation of the angular dependency is independent of the computation of the strengths of the angular channels $C_6^{(l_i, j_i)}$, these two tasks can be separated computationally which gives a significant advantage over existing implementations such as ARC \cite{Sibalic2017}. These existing implementations need to re-run their full calculations for every angle $\theta$ that one might be interested in, while we need to run the perturbative calculations once to obtain the channel strengths and can then reconstruct the angular dependencies within fractions of a second - even for as many pair states as shown in Figure \ref{fig:structuresInC6} simultaneously. This approach therefore makes searching the full manifold of $(n_1,\ n_2)$ Rydberg states at different angles feasible and allows to easily find pair states that match experimental requirements. In addition, the angular channel approach provides an intuitive and detailed understanding of the origin of the angular dependency and the implications of F\"orster defects. In order to make the computation speedup available to other users, we have integrated the angular channel code in the existing ARC Python package, which is available online: \cite{ARC}.

Our work makes a case for considering states with $n_1 \neq n_2$ for future experiments in order to access and utilise the wealth of Rydberg pair interaction potentials naturally available. Interaction potentials with $n_1 \neq n_2$ provide a much larger pool of pair states that might naturally implement a given target Hamiltonian. This may, in some cases, remove the need to construct these target Hamiltonians via techniques such as Hamiltonian engineering and therefore reduce experimental complexity. It also gives access to a large range of pair states that might fit other experimental constraints which $n_1 = n_2$ states cannot necessarily meet. Additionally, the insight from this approach can be used for various other applications such as e.g. interaction switching.

The presented approach was demonstrated for degenerate $m_j$ states, i.e. for the case without external electric or magnetic fields applied to the system. Applying electromagnetic fields allows to shift individual intermediate pair states into, or out of, resonance and therefore alter the interaction potential. It also induces an $m_j$-dependency to the energy defect such that the separation of sums in equation \ref{eqn:angularChannels} no longer holds true. The angular channel approach therefore does not hold in this regime. Is valid in the vdW limit, i.e. for interatomic distances larger than the vdW radius $r_{vdW}$ where second-order perturbation theory applies.

\begin{acknowledgments}
The authors acknowledge useful discussions with and feedback from Aaron Reinhard and Nikola Šibalić. We also thank Liam Gallagher for careful reading of the manuscript.

Financial support was provided by EPSRC  grant EP/V030280/1 (“Quantum optics using Rydberg polaritons”) and Durham University.
\end{acknowledgments}

\section*{Data Availability}
The angular channel code is integrated in ARC \cite{ARC}, and the preaclculated data for Rb-Rb and Cs-Cs interactions are available on Zenodo \cite{Zenodo} and via ARC \cite{ARC}.


\clearpage
\pagebreak
\onecolumngrid
\appendix

\section{Expressions for Radial and Angular Coupling Strengths}
\label{app:radialAngularExpressions}

\noindent The radial $\mathcal{R}$ and angular $\mathcal{D}^\alpha$ parts 
\[ d^\alpha = \braket{n^\prime, l^\prime, j^\prime, m_j^\prime|\hat{d}^\alpha|n, l, j, m_j} = \mathcal{R}(n, l, j;\ n^\prime, l^\prime, j^\prime) \mathcal{D}^\alpha(l, j, m_j;\ l^\prime, j^\prime, m_j^\prime) \]
of the dipole matrix element of a single-atom transition are given by \cite{Sibalic2017, Weber2017, Walker2008}
\begin{equation}
\label{eqn:radial}
\mathcal{R}(n, l;\ n^\prime, l^\prime) = (-1)^{l^\prime} \sqrt{(2l+1)(2l^\prime +1)} \begin{pmatrix} l & 1 & l^\prime \\ 0 & 0 & 0 \end{pmatrix} \ \int_0^\infty R_{nl}(r) er R_{n^\prime l^\prime}(r)\ r^2 dr
\end{equation}
and
\begin{align}
\label{eqn:angular} \notag
\mathcal{D}^\alpha(l, j, m_j;\ l^\prime, j^\prime, m_j^\prime) = &\ (-1)^{l+j+j^\prime+s-m_j+1} \sqrt{(2j+1)(2j^\prime +1)} \\
&\ \times  \begin{Bmatrix} j & 1 & j^\prime \\ l^\prime & s & l \end{Bmatrix} \begin{pmatrix}j & 1 & j^\prime \\ -m_j & -\alpha & m_j^\prime\end{pmatrix} .
\end{align}
$(:::)$ denotes the Wigner-$3\mathrm{j}$ symbol and $\{:::\}$ the Wigner-$6\mathrm{j}$ symbol. $s$ is the electron spin.

\section{Angular Momentum Channel Coupling Strength Evaluated at $\mathbf{\theta = 0}$}
\label{app:secondOrderAtThetaZero}

\begin{figure}
\centering
\includegraphics[width=\linewidth]{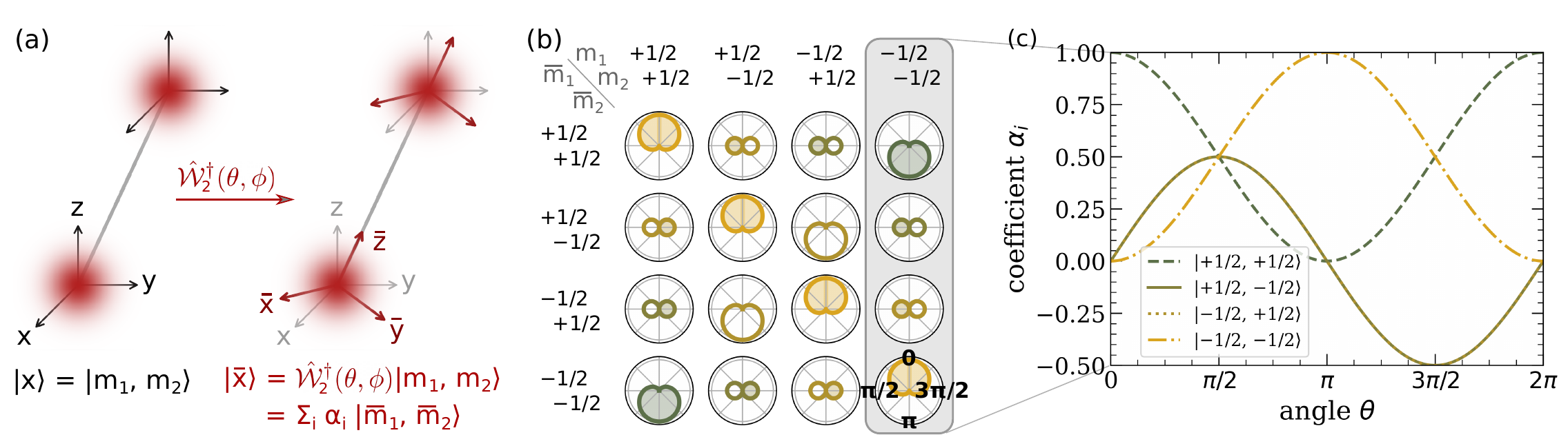}
\caption{\textbf{Wigner D basis rotations for $\mathbf{j_1 = j_2 = 1/2}$.} (a) Schematic representation of the basis rotation performed by application of the Wigner D-matrix. The initial state $\ket{m_1, m_2}$ is projected onto the rotated coordinate system where $\theta=0$ and $\ket{m_1, m_2} \to \sum_i \alpha_i \ket{\bar{m}_1, \bar{m}_2}$. (b) Projection of the total angular momentum states for varying angle $\theta$ and $j_1 = j_2 = 1/2$, i.e. $\alpha_i (\theta, \phi) = \braket{m_1^\prime, m_2^\prime |\hat{\mathcal{W}}_2^\dagger (\theta, \phi)| m_1, m_2}$ at $\phi=0$. Solid shaded areas indicate positive values. (c) Plot of $m_j$ coupling strength from $\ket{-1/2, -1/2} \to \sum_i \alpha_i \ket{\pm 1/2, \pm 1/2}$ for $\phi = 0$. Note that the detection probability in the new basis $|\alpha_i|^2$ is independent of $\phi$ and therefore determined by $\theta$ only.}
\label{fig:WignerD}
\end{figure}

\noindent To sum over all possible intermediate magnetic fine-structure states $(m_1^\prime, m_2^\prime)$ in equation \eqref{eqn:angularChannels} is tedious. To simplify numerical implementations of equation \eqref{eqn:angularChannels} one rotates the system such that $\theta^\prime = 0$ because then only the $\Delta M = 0$ processes couple, see also equation \eqref{eqn:dipoleInteractionSpherical} and Figure \ref{fig:dipoleSpherical} (a). To perform a rotation of the orbital angular momentum basis from coordinate system $X$ to $\bar{X}$, one applies Wigner D-matrices $\hat{\mathcal{W}}(\theta, \phi)$ such that $\ket{\bar{x}} = \hat{\mathcal{W}}^\dagger(\theta, \phi) \ket{x}$ \cite{Walker2008}, as shown in Figure \ref{fig:WignerD} (a). These basis rotations change the representation of the projection of the total orbital angular momentum $j$ from the $\{m_j\}$ basis to the $\{\bar{m}_j\}$ basis, but do not change any of the other quantum numbers $n$, $l$, or $j$. $\hat{\mathcal{W}}$ therefore depends on $j$, $m_j$ and $\bar{m}_j$, as well as on $\theta$ and $\phi$. The angle $\phi$, however, only contributes a complex phase factor $\pm \exp (\mp i n \phi)$ proportional to the change in total angular momentum, i.e. $n= \Delta m_1 + \Delta m_2$. The numerical implementation then reduces to
\[ C_\mathrm{6}(\theta, \phi) = \displaystyle\sum_{ \{ (l^\prime_i j^\prime_i) \} } C_6^{(l^\prime_i j^\prime_i)} \bra{f}\hat{\mathcal{W}}_2(\theta, \phi)\ \hat{\mathcal{D}}(l^\prime_i, j^\prime_i)\ \hat{\mathcal{W}}_2^\dagger(\theta, \phi) \ket{i} \]
where $\hat{\mathcal{D}}(l^\prime_i, j^\prime_i)$ is evaluated at $\theta = 0$ for every angular momentum channel $(l_1^\prime, j_1^\prime;\ l_2^\prime, j_2^\prime)$. Note that the Wigner-D matrix $\hat{\mathcal{W}}_2 = \hat{\mathcal{W}}_{atom\ 1} \otimes \hat{\mathcal{W}}_{atom\ 2}$ here constitutes of a product of the two single-atom Wigner-D matrices $\hat{\mathcal{W}}_{atom\ i}$.

The factor $\bra{f}\hat{\mathcal{W}}_2(\theta, \phi)\ \hat{\mathcal{D}}(l^\prime_i, j^\prime_i)\ \hat{\mathcal{W}}_2^\dagger(\theta, \phi) \ket{i}$ can be evaluated by implementing the Wigner D-matrix basis rotation and use the states expressed in the rotated basis to caculate
\begin{align} \small \notag
&\hspace{-2cm} \braket{\tilde{l}_1 \tilde{j}_1 \tilde{m}_{j1},\ \tilde{l}_2 \tilde{j}_2 \tilde{m}_{j2} | \hat{\mathcal{D}}(l^\prime_i j^\prime_i) | l_1 j_1 m_{j1},\ l_2 j_2 m_{j2} } \\  \notag
= & \displaystyle\sum_{\{ ( m^\prime_{j1}, m^\prime_{j2} ) \}} 
\left(\displaystyle\sum_{ \alpha\in\{0, \pm 1\} } \mathcal{C}(\alpha) \mathcal{D}^\alpha(l_1 j_1 m_{j1},\ l_1^\prime j_1^\prime m_{j1}^\prime) \mathcal{D}^{-\alpha}(l_2 j_2 m_{j2},\ l_2^\prime j_2^\prime m_{j2}^\prime)\right) \times \\
& \hspace{1.68cm} \left(\displaystyle\sum_{ \alpha^\prime\in\{0, \pm 1\} } \mathcal{C}(\alpha^\prime) \mathcal{D}^{\alpha^\prime} (l_1^\prime j_1^\prime m_{j1}^\prime,\ \tilde{l}_1 \tilde{j}_1 \tilde{m}_{j1} ) \mathcal{D}^{-\alpha^\prime}(l_2^\prime, j_2^\prime m_{j2}^\prime,\ \tilde{l}_2 \tilde{j}_2 \tilde{m}_{j2}) \right).
\end{align}
The different weights of the three different $m_j$ channels in $\hat{V}_{dd}(\theta=0)$ are implemented via
\[ \mathcal{C}(\alpha) =  \begin{cases} -2, & \alpha = 0 \\ -1, & \alpha = \pm 1 \end{cases}. \]

\section{Cancelling of Angular Dependencies For Different Channels}
\label{app:angularChannelCancelling}

The angular dependencies of channels can cancel in very specific circumstances. Since the angular dependency of the channels depends on the initial and final angular momentum projections $m_i$ and $\tilde{m}_i$, one has to study the relevant contributions. E.g. in the case for $\ket{S_{1/2}, S_{1/2}}$ states one finds that the angular dependencies of the $\ket{P_{1/2}, P_{1/2}}$ and the $\ket{P_{1/2}, P_{3/2}}$ (or $\ket{P_{3/2}, P_{1/2}}$) channels cancel for $(\tilde{m}_1, \tilde{m}_2) = (m_1, m_2)$. The same occurs for the $\ket{P_{1/2}, P_{1/2}}$ state and the intermediate states $\ket{S_{1/2}, S_{1/2}}$ and $\ket{S_{1/2}, D_{3/2}}$ (or $\ket{D_{3/2}, S_{1/2}}$), also with $(m_1^\prime, m_2^\prime) = (m_1, m_2)$. However, for the overall $C_6$ interaction to be isotropic, it is not sufficient that pairs of channels have canceling angular dependencies but also the respective channel coefficients have to be of very similar values. Away from F\"orster resonances this is usually given - but on F\"orster resonances of an angular channel it is not, as one can see in Figure \ref{fig:combinedKnowledge}. Away from F\"orster resonances there are many magnetic sub-state combinations for which the angular dependency of the different channels almost cancel to within a relative deviation of a few percent.

\section{Scaling of C6 with principal quantum number away from $n_1 = n_2$}
\label{app:scalingOfC6AwayFromN1equalsN2}

On the axis with $n_1 = n_2$, the magnitude of $C_6$ follows the well-known scaling law of $(n^\star)^{11}$. The scaling of the magnitude of $C_6$ away from the $n_1 = n_2$ axis, however, is a little more involved and depends on the ratio of $n_1$ and $n_2$. 

For illustration, we will have a look at the scaling of $C_6$ along a cross section where $n_2$ remains constant and $n_1$ varies, such as e.g. in Figure \ref{fig:combinedKnowledge}. The scaling of $C_6$ with $n_1$ behaves differently for lower $n_1$ than for higher $n_1$ relative to $n_2$. To understand why, we need to look at the different contributions in the equation to calculate $C_6$ perturbatively, i.e. equation \ref{eqn:angularChannels}. Here, we can ignore the angular dependency for now as it does not contribute to the scaling with $n^\star$
\[ C_6 \propto \displaystyle\sum_{ \{ (n^\prime_i l^\prime_i j^\prime_i) \} } \frac{\mathcal{R}_1 \mathcal{R}_2 \mathcal{R}_1^\prime \mathcal{R}_2^\prime}{\Delta(n^\prime_i, l^\prime_i, j^\prime_i)}\]
The energy defect $\Delta(n^\prime_i, l^\prime_i, j^\prime_i)= \Delta(n^\prime_1, l^\prime_1, j^\prime_1) + \Delta(n^\prime_2, l^\prime_2, j^\prime_2) = \Delta_1 + \Delta_2$ in the denominator consists of separate contributions from the two atoms, respectively. To study the $n^\star$ scaling away from F\"orster resonance, we can ignore the sum for a moment and then find that $C_6$ scales along the cross section as
\[ C_6 \propto \frac{\mathcal{R}_1 \mathcal{R}_2 \mathcal{R}_1^\prime \mathcal{R}_2^\prime}{\Delta_1+\Delta_2} \propto  \frac{(n_1^\star)^4}{(n_1^\star)^{-3}+\Delta_2} \]
As $\Delta_2$ remains constant along the cross section, the scaling of the denominator is not constant but varies with $n_1$ for any given $n_2$. For $n_1 > n_2$, the denominator is dominated by the energy defect contribution $\Delta_2 \propto (n_2^\star)^{-3}$, such that the overall scaling of $C_6$ approaches $\propto (n_1^\star)^4$. Whereas for $n_1 < n_2$ the energy defect contribution is dominated by $\Delta_1$, such that the overall scaling approaches $\propto (n_1^\star)^7$ in that regime.

\section{F\"orster Resonance Lines For Angular Momentum Channels Of Different Initial States}
\label{app:FoersterResonanceLines}

In the following, we list the F\"orster resonance lines of relevance for various pair states. The resonances were obtained for rubidium but are essentially similar for cesium. The Bohr model is sufficient to calculate the approximate position of the resonances since more elaborate energy level structure models like e.g. the Sommerfeld model do not lead to changes in the state energies large enough to shift the F\"orster resonance lines significantly.

Note that the symmetric channels with $(l_i, j_i) = (l_j, j_j)$ produce the same behaviour for exchange of $n_i$ and $n_j$, while the asymmetric channels do not. This is indicated by the different indices $i,j$ (symmetric) versus $1,2$ (asymmetric) for the $\Delta n_x$ parameters in the following lists.

\subsection{$\mathbf{\ket{n_1 S_{1/2}, n_2 S_{1/2}} \leftrightarrow \ket{(n_1 + \Delta n_1) P_{J_1^\prime}, (n_2 + \Delta n_2)P_{J_2^\prime}}}$}

$\ket{(n_1 + \Delta n_1) P_{J_1^\prime}, (n_2 + \Delta n_2) P_{J_2^\prime}}$: $\Delta n_i \leq -1, \Delta n_j \geq 0$ with $i \neq j$

\subsection{$\mathbf{\ket{n_1 P_{J}, n_2 P_{J}} \leftrightarrow \ket{(n_1 + \Delta n_1) (L_1 \pm 1)_{J_1^\prime}, (n_2 + \Delta n_2)(L_2 \pm 1)_{J_2^\prime}}}$ with $\mathbf{J_i \in \{1/2, 3/2\}}$}

Check which $J^\prime$ values are coupled to $nP_{1/2}$ states.

$\ket{(n_1 + \Delta n_1) S_{1/2}, (n_2 + \Delta n_2) S_{1/2}}$: $\Delta n_i \leq 0, \Delta n_j \geq 1$

$\ket{(n_1 + \Delta n_1) S_{1/2}, (n_2 + \Delta n_2) D_{J_2^\prime}}$: $(\Delta n_1 \leq 0, \Delta n_2 \geq -1)$ and $(\Delta n_1 \geq 1, \Delta n_2 \leq -2)$

$\ket{(n_1 + \Delta n_1) D_{J^\prime}, (n_2 + \Delta n_2) D_{J^\prime}}$: $\Delta n_i \leq -2, \Delta n_j \geq -1$ for $i \neq j$ and $J^\prime \in {3/2, 5/2}$

$\ket{(n_1 + \Delta n_1) D_{3/2}, (n_2 + \Delta n_2) D_{5/2}}$: $(\Delta n_1 \leq -2, \Delta n_2 \geq -1)$ and $(\Delta n_1 \geq -1, \Delta n_2 \leq -2)$

\subsection{$\mathbf{\ket{n_1 D_{J}, n_2 D_{J}} \leftrightarrow \ket{(n_1 + \Delta n_1) (L_1 \pm 1)_{J_1^\prime}, (n_2 + \Delta n_2)(L_2 \pm 1)_{J_2^\prime}}}$ with $\mathbf{J \in \{3/2, 5/2\}}$}

Check which $J^\prime$ values are coupled to $nD_{3/2}$ and which to $nD_{5/2}$ states.

$\ket{(n_1 + \Delta n_1) P_{J^\prime}, (n_2 + \Delta n_2) P_{J^\prime}}$: $\Delta n_i \leq 1, \Delta n_j \geq 2$ with $i \neq j$

$\ket{(n_1 + \Delta n_1) P_{1/2}, (n_2 + \Delta n_2) P_{3/2}}$: $(\Delta n_1 \leq 1, \Delta n_2 \geq 2)$ and $(\Delta n_1 \geq 2, \Delta n_2 \leq 1)$

$\ket{(n_1 + \Delta n_1) P_{J_1^\prime}, (n_2 + \Delta n_2) F_{J_2^\prime}}$: $(\Delta n_1 \leq 1, \Delta n_2 \geq -1)$ and $(\Delta n_1 \geq 2, \Delta n_2 \leq -2)$ with $J_1^\prime \in \{1/2, 3/2\}$ and $J_1^\prime \in \{5/2, 7/2\}$

$\ket{(n_1 + \Delta n_1) F_{J^\prime}, (n_2 + \Delta n_2) F_{J^\prime}}$: $\Delta n_i \leq -2, \Delta n_j \geq -1$ with $i \neq j$ and $J^\prime \in \{5/2, 7/2\}$

$\ket{(n_1 + \Delta n_1) F_{5/2}, (n_2 + \Delta n_2) F_{7/2}}$: $(\Delta n_1 \leq -2, \Delta n_2 \geq -1)$ and $(\Delta n_1 \geq -1, \Delta n_2 \leq -2)$

\subsection{$\mathbf{\ket{n_1 F_{J}, n_2 F_{J}} \leftrightarrow \ket{(n_1 + \Delta n_1) (L_1 \pm 1)_{J_1^\prime}, (n_2 + \Delta n_2)(L_2 \pm 1)_{J_2^\prime}}}$ with $\mathbf{J \in \{5/2, 7/2\}}$}

Check which $J^\prime$ values are coupled to $nF_{5/2}$ and which to $nF_{7/2}$ states.

$\ket{(n_1 + \Delta n_1) D_{J^\prime}, (n_2 + \Delta n_2) D_{J^\prime}}$: $\Delta n_i \leq 1, \Delta n_j \geq 2$ with $i \neq j$

$\ket{(n_1 + \Delta n_1) D_{3/2}, (n_2 + \Delta n_2) D_{5/2}}$: $(\Delta n_1 \leq 1, \Delta n_2 \geq 2)$ and $(\Delta n_1 \geq 2, \Delta n_2 \leq 1)$

$\ket{(n_1 + \Delta n_1) D_{J_1^\prime}, (n_2 + \Delta n_2) G_{J_2^\prime}}$: $(\Delta n_1 \leq 1, \Delta n_2 \geq 0)$ and $(\Delta n_1 \geq 2, \Delta n_2 \leq -1)$ with $J_1^\prime \in \{3/2, 5/2\}$ and $J_1^\prime \in \{7/2, 9/2\}$

$\ket{(n_1 + \Delta n_1) G_{J^\prime}, (n_2 + \Delta n_2) G_{J^\prime}}$: $\Delta n_i \leq -1, \Delta n_j \geq 1$ with $i \neq j$ and $J^\prime \in \{7/2, 9/2\}$

$\ket{(n_1 + \Delta n_1) G_{7/2}, (n_2 + \Delta n_2) G_{9/2}}$: $(\Delta n_1 \leq -1, \Delta n_2 \geq 1)$ and $(\Delta n_1 \geq 1, \Delta n_2 \leq -1)$

\subsection{$\mathbf{\ket{n_1 S_{1/2}, n_2 D_{J_2}} \leftrightarrow \ket{(n_1 + \Delta n_1) P_{J_1^\prime}, (n_2 + \Delta n_2)(L_2 \pm 1)_{J_2^\prime}}}$ with $\mathbf{J_2 \in \{3/2, 5/2\}}$}

Check which $J_2^\prime$ values are coupled to $nD_{3/2}$ and which to $nD_{5/2}$ states.

$\ket{(n_1 + \Delta n_1) P_{J_1^\prime}, (n_2 + \Delta n_2) P_{J_2^\prime}}$: $(\Delta n_1 \leq -1, \Delta n_2 \geq 2)$ and $(\Delta n_1 \geq 0, \Delta n_2 \leq 1)$ with $J_{1,2}^\prime \in \{1/2, 3/2\}$

$\ket{(n_1 + \Delta n_1) P_{J_1^\prime}, (n_2 + \Delta n_2) F_{J_2^\prime}}$: $(\Delta n_1 \leq -1, \Delta n_2 \geq -1)$ and $(\Delta n_1 \geq 0, \Delta n_2 \leq -2)$ with $J_1^\prime \in \{1/2, 3/2\}$ and $J_2^\prime \in \{5/2, 7/2\}$

\subsection{$\mathbf{\ket{n_1 S_{1/2}, n_2 F_{J_2}} \leftrightarrow \ket{(n_1 + \Delta n_1) P_{J_1^\prime}, (n_2 + \Delta n_2)(L_2 \pm 1)_{J_2^\prime}}}$ with $\mathbf{J_2 \in \{5/2, 7/2\}}$}

Check which $J_2^\prime$ values are coupled to $nF_{5/2}$ and which to $nF_{7/2}$ states.

$\ket{(n_1 + \Delta n_1) P_{J_1^\prime}, (n_2 + \Delta n_2) D_{J_2^\prime}}$: $(\Delta n_1 \leq -1, \Delta n_2 \geq 2)$ and $(\Delta n_1 \geq 0, \Delta n_2 \leq 1)$ with $J_{1}^\prime \in \{1/2, 3/2\}$ and $J_{2}^\prime \in \{3/2, 5/2\}$

$\ket{(n_1 + \Delta n_1) P_{J_1^\prime}, (n_2 + \Delta n_2) G_{J_2^\prime}}$: $(\Delta n_1 \leq -1, \Delta n_2 \geq 0)$ and $(\Delta n_1 \geq 0, \Delta n_2 \leq -1)$ with $J_1^\prime \in \{1/2, 3/2\}$ and $J_2^\prime \in \{7/2, 9/2\}$

\subsection{$\mathbf{\ket{n_1 P_{J_1}, n_2 F_{J_2}} \leftrightarrow \ket{(n_1 + \Delta n_1) (L_1 \pm 1)_{J_1^\prime}, (n_2 + \Delta n_2)(L_2 \pm 1)_{J_2^\prime}}}$ with $\mathbf{J_1 \in \{1/2, 3/2\}}$ and $\mathbf{J_2 \in \{5/2, 7/2\}}$}

Check which $J_1^\prime$ values are coupled to $nP_{1/2}$ and which to $nP_{3/2}$ states. Also check which $J_2^\prime$ values are coupled to $nF_{5/2}$ and which to $nF_{7/2}$ states.

$\ket{(n_1 + \Delta n_1) S_{1/2}, (n_2 + \Delta n_2) D_{J_2^\prime}}$: $(\Delta n_1 \leq 0, \Delta n_2 \geq 2)$ and $(\Delta n_1 \geq 1, \Delta n_2 \leq 1)$ with $J_{2}^\prime \in \{3/2, 5/2\}$

$\ket{(n_1 + \Delta n_1) S_{1/2}, (n_2 + \Delta n_2) G_{J_2^\prime}}$: $(\Delta n_1 \leq 0, \Delta n_2 \geq 0)$ and $(\Delta n_1 \geq 1, \Delta n_2 \leq -1)$ with $J_2^\prime \in \{7/2, 9/2\}$

$\ket{(n_1 + \Delta n_1) D_{J_1^\prime}, (n_2 + \Delta n_2) D_{J_2^\prime}}$: $(\Delta n_1 \leq -2, \Delta n_2 \geq 2)$ and $(\Delta n_1 \geq -1, \Delta n_2 \leq 1)$ with $J_{1,2}^\prime \in \{3/2, 5/2\}$

$\ket{(n_1 + \Delta n_1) D_{J_1^\prime}, (n_2 + \Delta n_2) G_{J_2^\prime}}$: $(\Delta n_1 \leq -2, \Delta n_2 \geq 1)$ and $(\Delta n_1 \geq -1, \Delta n_2 \leq -1)$ with $J_1^\prime \in \{1/2, 3/2\}$ and $J_2^\prime \in \{7/2, 9/2\}$

\section{2D maps for Rubidium pair states}
\label{app:otherStateMapsRb}

$C_6$ absolute value and sign maps for $\theta = 0, \pi / 2$  in rubidium are shown for $\ket{S_{1/2}, S_{1/2}}$, $\ket{P_{J_1}, P_{J_2}}$, $\ket{D_{J_1}, D_{J_2}}$, and $\ket{F_{J_1}, F_{J_2}}$ pair states. Additionally, $\ket{S_{1/2}, D_{J}}$ and $\ket{P_{J_1}, F_{J_2}}$ pair state maps are shown as they can also be coupled via 2- and 3-photon transitions from the ground state, respectively.

Pair states with strong angular dependencies between $\theta = 0, \pi/2$ can be identified by looking for bright/dark bands that shift position between the two angles (e.g. $\ket{n_1 F_J, n_2F_J}$), or dark bands that are present in just one of the two plots (see e.g. $\ket{n_1 S_{1/2}, n_2D_J}$). Additionally, a change in sign between the two angles also indicates a strong angular dependency plus an additional zero crossing of $C_6(\theta)$. Such pair states are present in every plot.

Furthermore, individual pair states can have much stronger $C_6$ values than the surrounding pair states. Usually, the F\"orster resonance lines carry such candidates. Some of these F\"orster resonant pairs possess strongly enhanced $C_6$ values if the energy defect is particularly small. Besides, one can find isolated resonances such as $\ket{57P_{3/2}, 82F_{7/2}}$ or $\ket{113S_{1/2}, 60D_{5/2}}$. All pair states which have a $C_6$ value that is at least by a factor of $\times 100$ stronger than the $C_6$ values in its vicinity are mentioned in the figure captions.

The 2D map plots for all 2- or 3-photon addressable pair states are included below since the pair states for different angular quantum number combinations produce different sets of interesting features that are relevant for different applications. However, as different experiments will seek different properties in the pair states, we cannot compile an exhaustive list of interesting states. The precalculated angular channel datasets and the angular channel code are available on GitHub \citep{ARC} so that users can run their own search for interesting pair states, e.g. with different $m_j$ values or for states matching particular experimental requirements.

\begin{figure}[H]
\centering
\includegraphics[width=\linewidth]{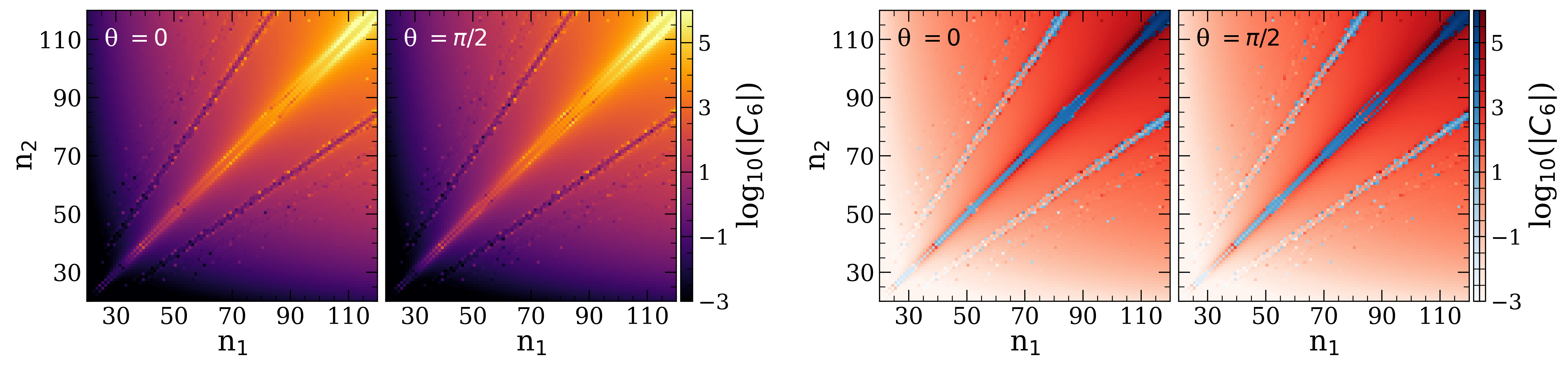}
\caption{\textbf{$\mathbf{C_6(n_1, n_2)}$ map for $\mathbf{\ket{n_1 S_{1/2},\ n_2 S_{1/2}}}$ pair states in rubidium.} Absolute value (left) and sign (right) maps with $C_6 < 0$: blue, $C_6 > 0$: red. $m_1 = \pm j_1,\ m_2 = \pm j_2$ and $\tilde{m}_i = m_i$. The $C_6$ value of (57,79) is a factor of 110 stronger than that of the surrounding pair states.}
\label{fig:RbS05RbS05}
\end{figure}

\begin{figure}[H]
\centering
\includegraphics[width=\linewidth]{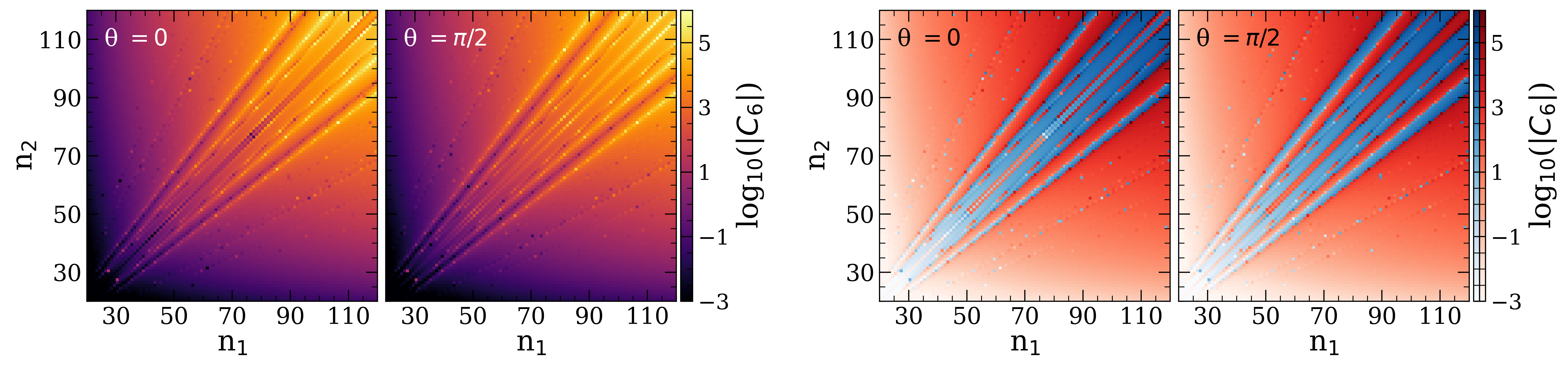}
\caption{\textbf{$\mathbf{C_6(n_1, n_2)}$ map for $\mathbf{\ket{n_1 P_{1/2},\ n_2 P_{1/2}}}$ pair states in rubidium.} Absolute value (left) and sign (right) maps with $C_6 < 0$: blue, $C_6 > 0$: red. $m_1 = \pm j_1,\ m_2 = \pm j_2$ and $\tilde{m}_i = m_i$. The state (27, 30) has a $C_6$ value which is at least a factor of 280 higher than of the surrounding pair states.}
\label{fig:RbP05RbP05}
\end{figure}

\begin{figure}[H]
\centering
\includegraphics[width=\linewidth]{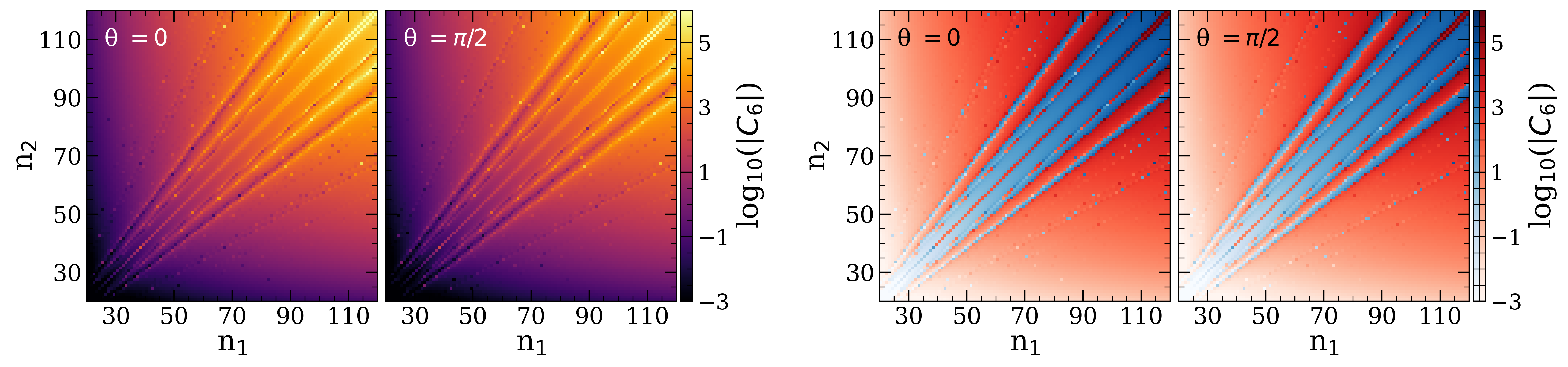}
\caption{\textbf{$\mathbf{C_6(n_1, n_2)}$ map for $\mathbf{\ket{n_1 P_{3/2},\ n_2 P_{3/2}}}$ pair states in rubidium.} Absolute value (left) and sign (right) maps with $C_6 < 0$: blue, $C_6 > 0$: red. $m_1 = \pm j_1,\ m_2 = \pm j_2$ and $\tilde{m}_i = m_i$.}
\label{fig:RbP15RbP15}
\end{figure}

\begin{figure}[H]
\centering
\includegraphics[width=\linewidth]{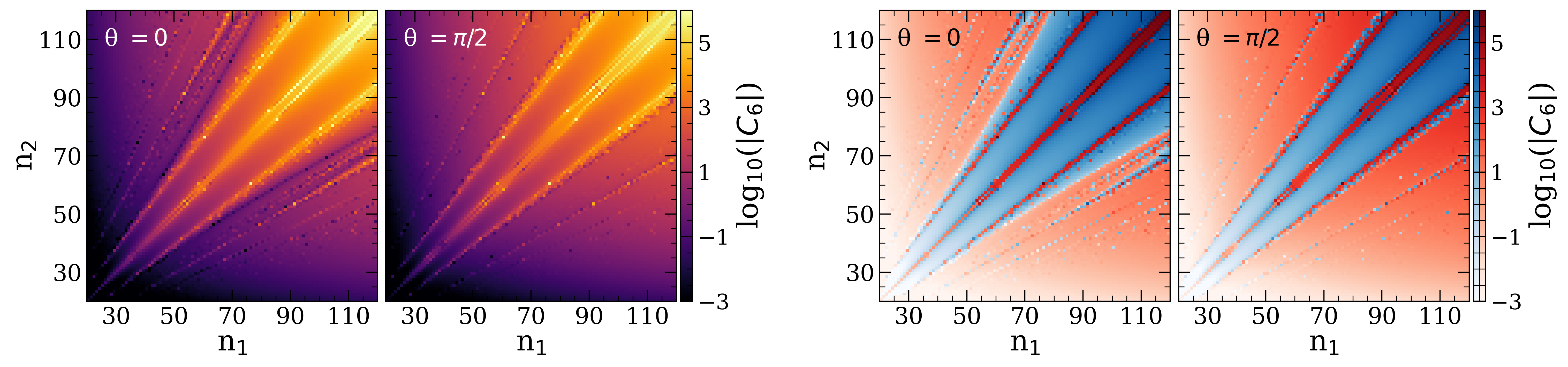}
\caption{\textbf{$\mathbf{C_6(n_1, n_2)}$ map for $\mathbf{\ket{n_1 D_{3/2},\ n_2 D_{3/2}}}$ pair states in rubidium.} Absolute value (left) and sign (right) maps with $C_6 < 0$: blue, $C_6 > 0$: red. $m_1 = \pm j_1,\ m_2 = \pm j_2$ and $\tilde{m}_i = m_i$. The state (53, 91) has a $C_6$ value which is at least a factor of 534 higher than of the surrounding pair states, and the factor is just slightly less with 523 for the (60, 76) pair state.}
\label{fig:RbD15RbD15}
\end{figure}

\begin{figure}[H]
\centering
\includegraphics[width=\linewidth]{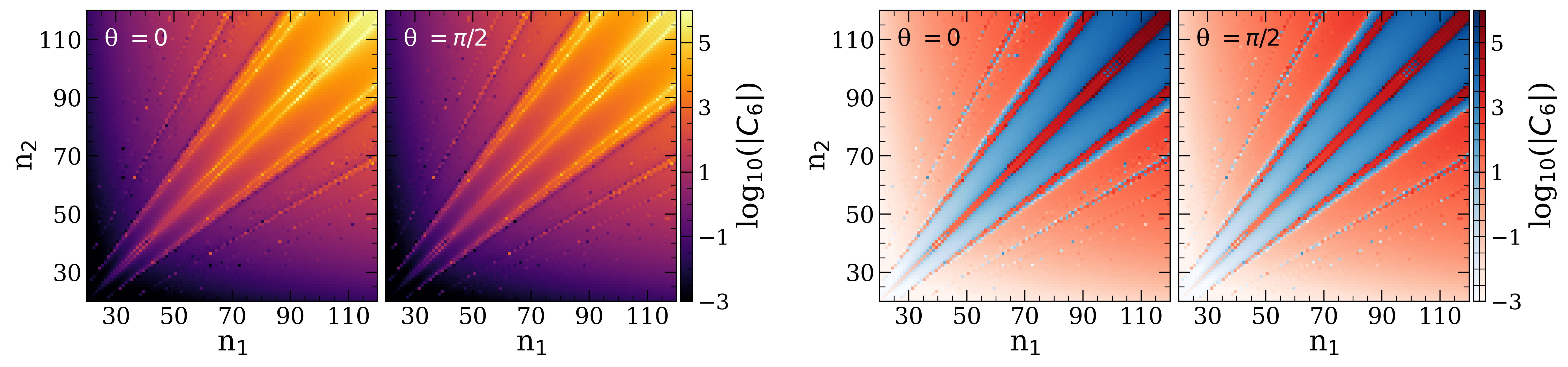}
\caption{\textbf{$\mathbf{C_6(n_1, n_2)}$ map for $\mathbf{\ket{n_1 D_{5/2},\ n_2 D_{5/2}}}$ pair states in rubidium.} Absolute value (left) and sign (right) maps with $C_6 < 0$: blue, $C_6 > 0$: red. $m_1 = \pm j_1,\ m_2 = \pm j_2$ and $\tilde{m}_i = m_i$. The state (36, 62) has a $C_6$ value which is at least a factor of 551 higher than of the surrounding pair states.}
\label{fig:RbD25RbD25}
\end{figure}

\begin{figure}[H]
\centering
\includegraphics[width=\linewidth]{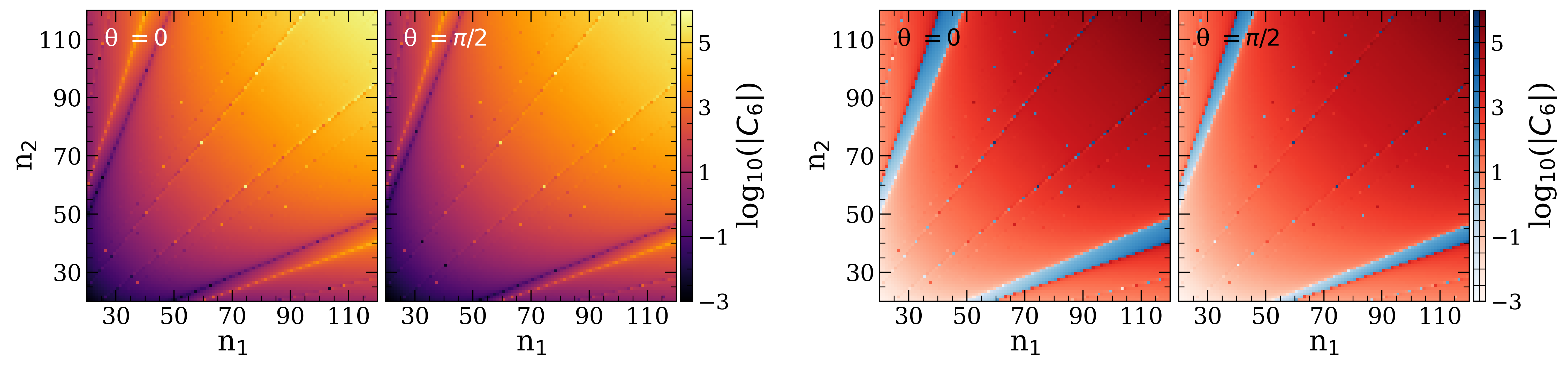}
\caption{\textbf{$\mathbf{C_6(n_1, n_2)}$ map for $\mathbf{\ket{n_1 F_{5/2},\ n_2 F_{5/2}}}$ pair states in rubidium.} Absolute value (left) and sign (right) maps with $C_6 < 0$: blue, $C_6 > 0$: red. $m_1 = \pm j_1,\ m_2 = \pm j_2$ and $\tilde{m}_i = m_i$.}
\label{fig:RbF25RbF25}
\end{figure}

\begin{figure}[H]
\centering
\includegraphics[width=\linewidth]{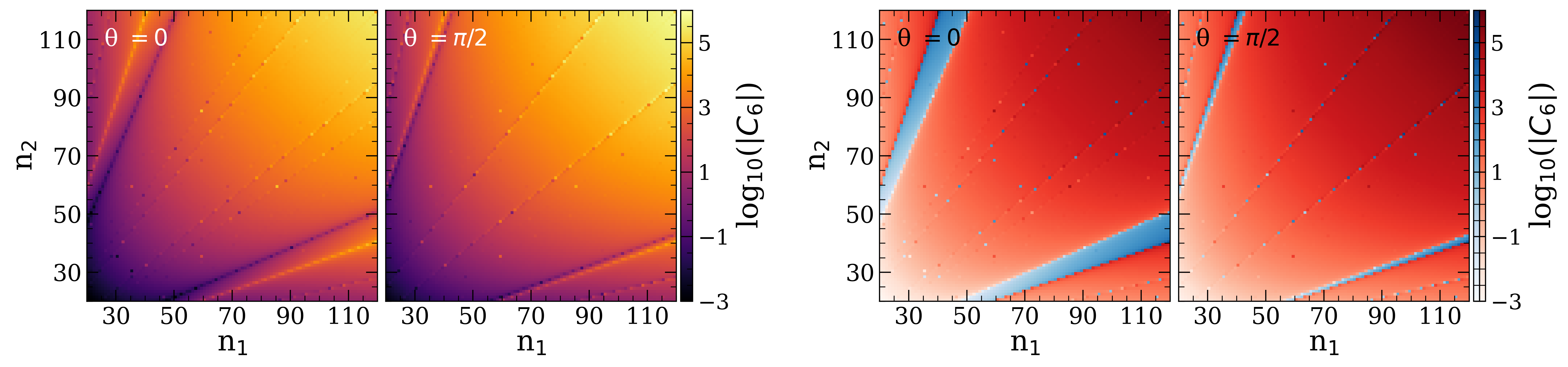}
\caption{\textbf{$\mathbf{C_6(n_1, n_2)}$ map for $\mathbf{\ket{n_1 F_{7/2},\ n_2 F_{7/2}}}$ pair states in rubidium.} Absolute value (left) and sign (right) maps with $C_6 < 0$: blue, $C_6 > 0$: red. $m_1 = \pm j_1,\ m_2 = \pm j_2$ and $\tilde{m}_i = m_i$.}
\label{fig:RbF35RbF35}
\end{figure}

\begin{figure}[H]
\centering
\includegraphics[width=\linewidth]{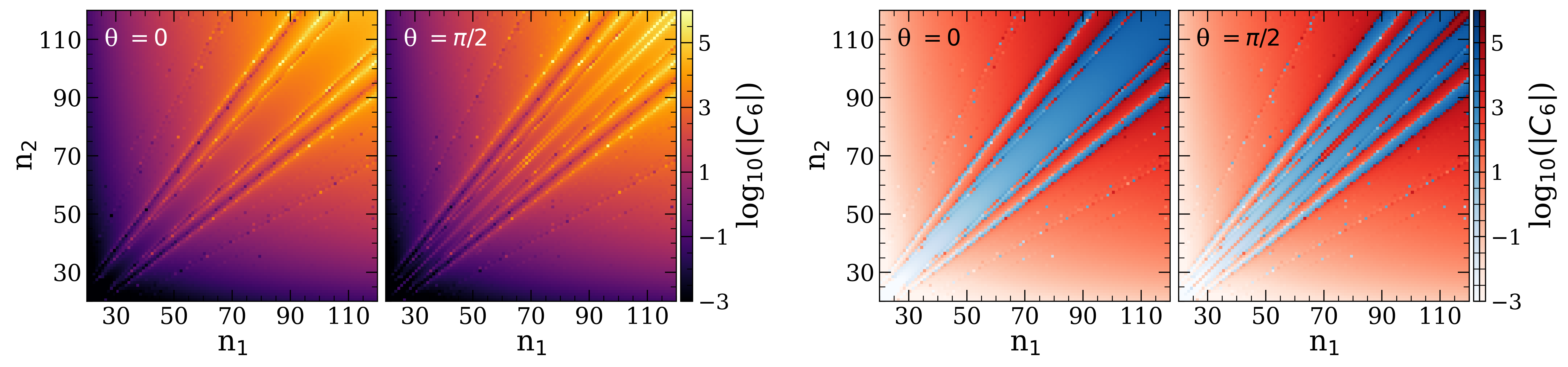}
\caption{\textbf{$\mathbf{C_6(n_1, n_2)}$ map for $\mathbf{\ket{n_1 P_{1/2},\ n_2 P_{3/2}}}$ pair states in rubidium.} Absolute value (left) and sign (right) maps with $C_6 < 0$: blue, $C_6 > 0$: red. $m_1 = \pm j_1,\ m_2 = \pm j_2$ and $\tilde{m}_i = m_i$. The state (106, 80) has a $C_6$ value which is at least a factor of 176 higher than of the surrounding pair states.}
\label{fig:RbP05RbP15}
\end{figure}

\begin{figure}[H]
\centering
\includegraphics[width=\linewidth]{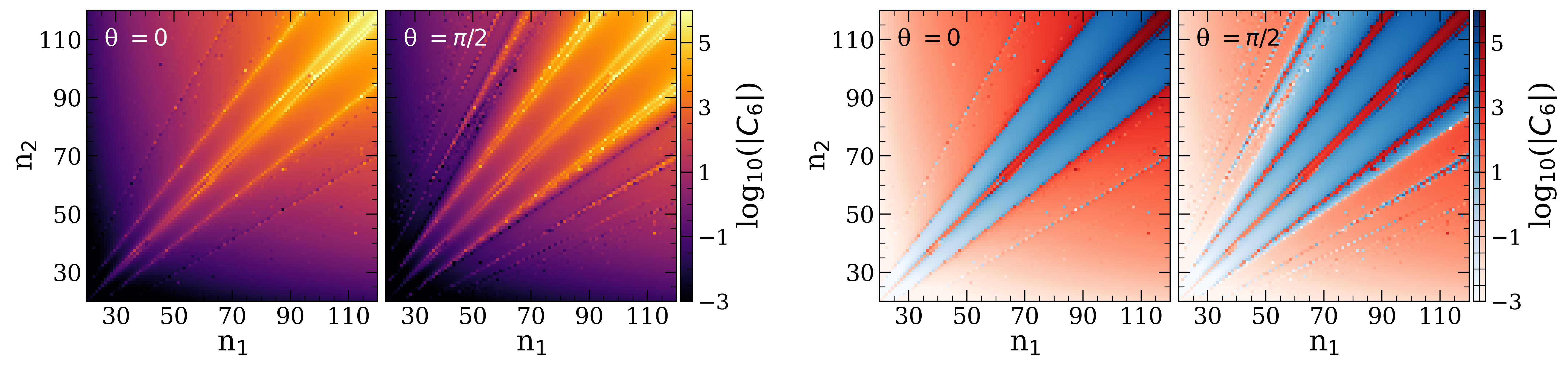}
\caption{\textbf{$\mathbf{C_6(n_1, n_2)}$ map for $\mathbf{\ket{n_1 D_{3/2},\ n_2 D_{5/2}}}$ pair states in rubidium.} Absolute value (left) and sign (right) maps with $C_6 < 0$: blue, $C_6 > 0$: red. $m_1 = \pm j_1,\ m_2 = \pm j_2$ and $\tilde{m}_i = m_i$. The state (43, 112) has a $C_6$ value which is at least a factor of 124 higher than of the surrounding pair states.}
\label{fig:RbD15RbD25}
\end{figure}

\begin{figure}[H]
\centering
\includegraphics[width=\linewidth]{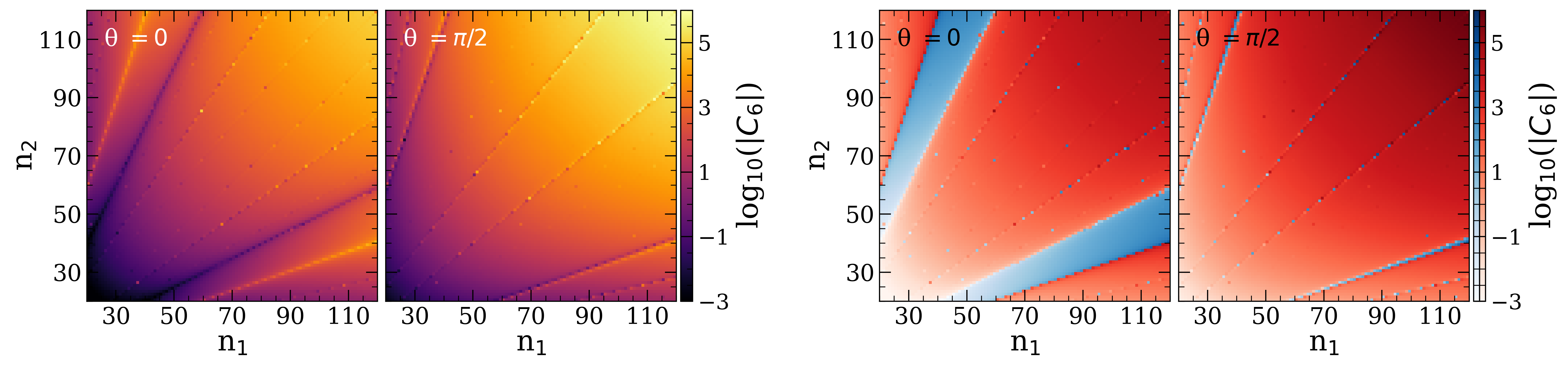}
\caption{\textbf{$\mathbf{C_6(n_1, n_2)}$ map for $\mathbf{\ket{n_1 F_{5/2},\ n_2 F_{7/2}}}$ pair states in rubidium.} Absolute value (left) and sign (right) maps with $C_6 < 0$: blue, $C_6 > 0$: red. $m_1 = \pm j_1,\ m_2 = \pm j_2$ and $\tilde{m}_i = m_i$}
\label{fig:RbF25RbF35}
\end{figure}

\begin{figure}[H]
\centering
\includegraphics[width=\linewidth]{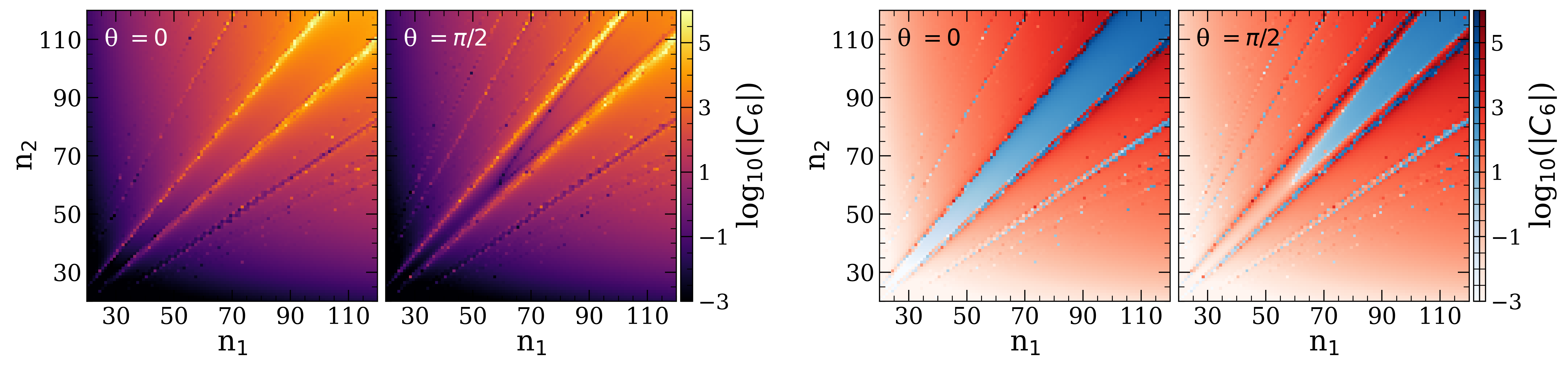}
\caption{\textbf{$\mathbf{C_6(n_1, n_2)}$ map for $\mathbf{\ket{n_1 S_{1/2},\ n_2 D_{3/2}}}$ pair states in rubidium.} Absolute value (left) and sign (right) maps with $C_6 < 0$: blue, $C_6 > 0$: red. $m_1 = \pm j_1,\ m_2 = \pm j_2$ and $\tilde{m}_i = m_i$. The state (52, 73) has a $C_6$ value which is at least a factor of 164 higher than of the surrounding pair states.}
\label{fig:RbS05RbD15}
\end{figure}

\begin{figure}[H]
\centering
\includegraphics[width=\linewidth]{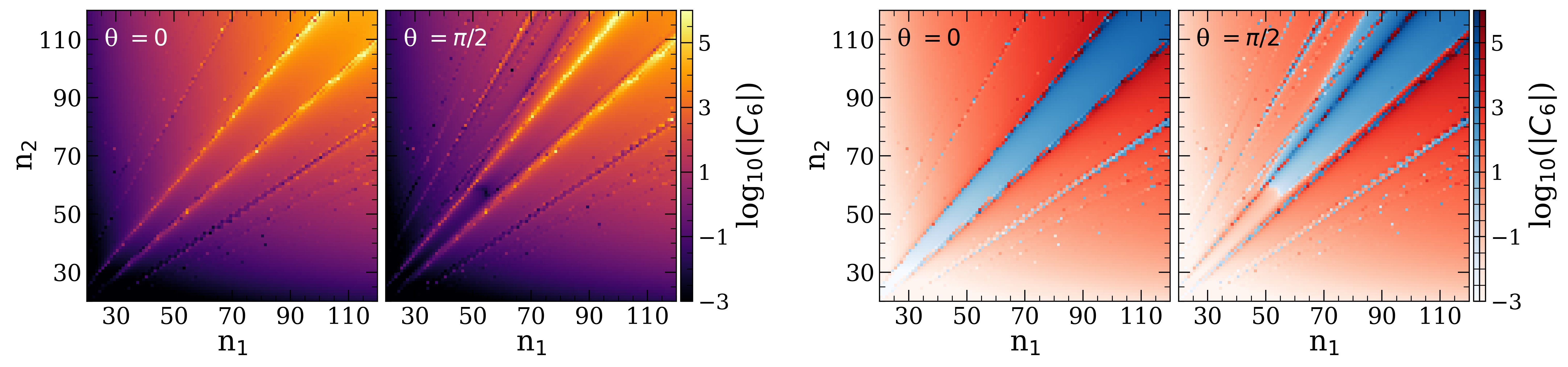}
\caption{\textbf{$\mathbf{C_6(n_1, n_2)}$ map for $\mathbf{\ket{n_1 S_{1/2},\ n_2 D_{5/2}}}$ pair states in rubidium.} Absolute value (left) and sign (right) maps with $C_6 < 0$: blue, $C_6 > 0$: red. $m_1 = \pm j_1,\ m_2 = \pm j_2$ and $\tilde{m}_i = m_i$. The state (71, 78) has a $C_6$ value which is at least a factor of 212 higher than of the surrounding pair states and the pair state (113, 66) has a minimum ratio of 118.}
\label{fig:RbS05RbD25}
\end{figure}

\begin{figure}[H]
\centering
\includegraphics[width=\linewidth]{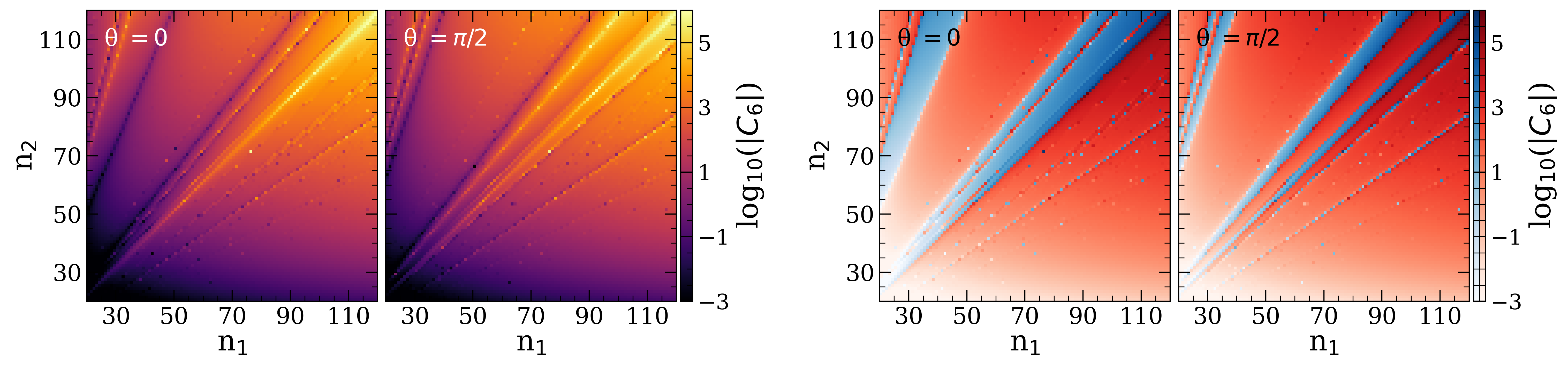}
\caption{\textbf{$\mathbf{C_6(n_1, n_2)}$ map for $\mathbf{\ket{n_1 P_{1/2},\ n_2 F_{5/2}}}$ pair states in rubidium.} Absolute value (left) and sign (right) maps with $C_6 < 0$: blue, $C_6 > 0$: red. $m_1 = \pm j_1,\ m_2 = \pm j_2$ and $\tilde{m}_i = m_i$. The state (55, 78) has a $C_6$ value which is at least a factor of 169 higher than of the surrounding pair states.}
\label{fig:RbP05RbF25}
\end{figure}

\begin{figure}[H]
\centering
\includegraphics[width=\linewidth]{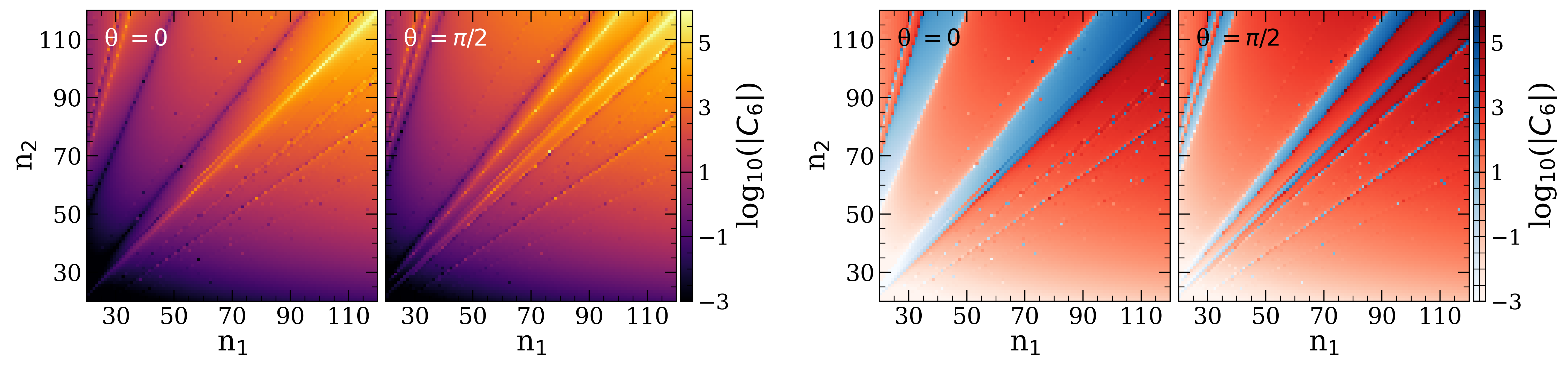}
\caption{\textbf{$\mathbf{C_6(n_1, n_2)}$ map for $\mathbf{\ket{n_1 P_{1/2},\ n_2 F_{7/2}}}$ pair states in rubidium.} Absolute value (left) and sign (right) maps with $C_6 < 0$: blue, $C_6 > 0$: red. $m_1 = \pm j_1,\ m_2 = \pm j_2$ and $\tilde{m}_i = m_i$. The state (55, 78) has a $C_6$ value which is at least a factor of 127 higher than of the surrounding pair states.}
\label{fig:RbP05RbF35}
\end{figure}

\begin{figure}[H]
\centering
\includegraphics[width=\linewidth]{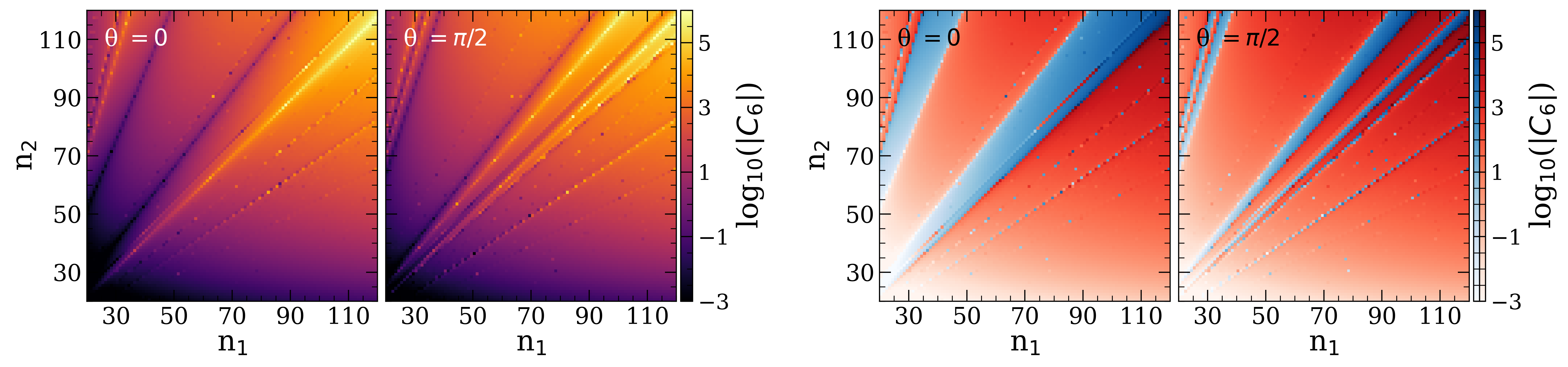}
\caption{\textbf{$\mathbf{C_6(n_1, n_2)}$ map for $\mathbf{\ket{n_1 P_{3/2},\ n_2 F_{5/2}}}$ pair states in rubidium.} Absolute value (left) and sign (right) maps with $C_6 < 0$: blue, $C_6 > 0$: red. $m_1 = \pm j_1,\ m_2 = \pm j_2$ and $\tilde{m}_i = m_i$. The state (57, 82) has a $C_6$ value which is at least a factor of 263 higher than of the surrounding pair states.}
\label{fig:RbP15RbF25}
\end{figure}

\begin{figure}[H]
\centering
\includegraphics[width=\linewidth]{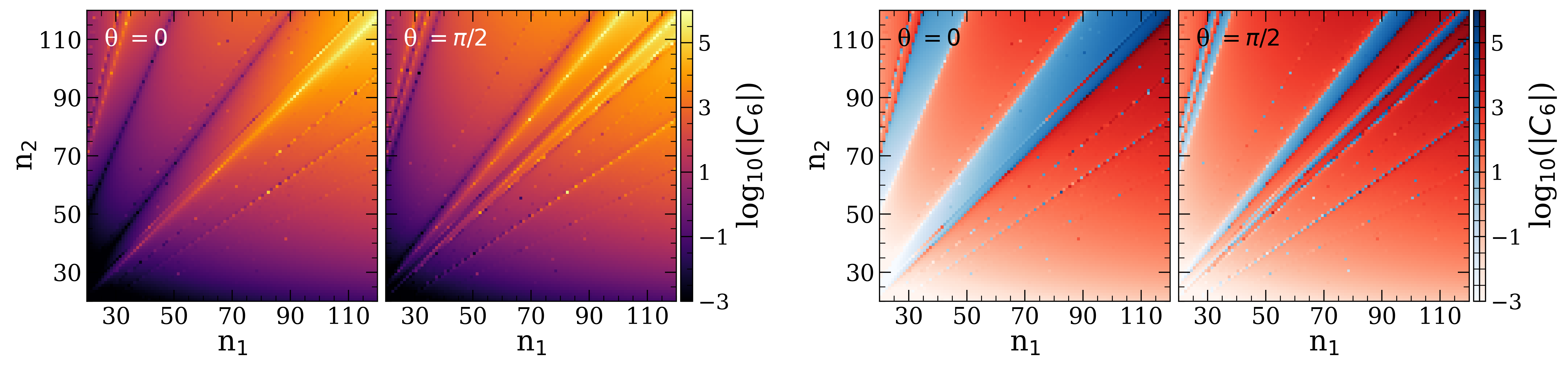}
\caption{\textbf{$\mathbf{C_6(n_1, n_2)}$ map for $\mathbf{\ket{n_1 P_{3/2},\ n_2 F_{7/2}}}$ pair states in rubidium.} Absolute value (left) and sign (right) maps with $C_6 < 0$: blue, $C_6 > 0$: red. $m_1 = \pm j_1,\ m_2 = \pm j_2$ and $\tilde{m}_i = m_i$. The state (57, 82) has a $C_6$ value which is at least a factor of 1282 higher than of the surrounding pair states, which is three orders of magnitude!}
\label{fig:RbP15RbF35}
\end{figure}

Note that \cref{fig:RbS05RbS05,fig:RbP05RbP05,fig:RbP15RbP15,fig:RbD15RbD15,fig:RbD25RbD25,fig:RbF25RbF25,fig:RbF35RbF35,fig:RbP05RbP15,fig:RbD15RbD25,fig:RbF25RbF35,fig:RbS05RbD15,fig:RbS05RbD25,fig:RbP05RbF25,fig:RbP05RbF35,fig:RbP15RbF25,fig:RbP15RbF35} show states with $m_i = \tilde{m_i}$ and $m_i =  j_i$, i.e. the states often chosen in experiments. These 2D maps look different for different $m_j$ states.


\section{2D maps for Cesium pair states}
\label{app:otherStateMapsCs}

$C_6$ absolute value and sign maps for $\theta = 0, \pi / 2$  in rubidium are shown for $\ket{S_{1/2}, S_{1/2}}$, $\ket{P_{J_1}, P_{J_2}}$, $\ket{D_{J_1}, D_{J_2}}$, and $\ket{F_{J_1}, F_{J_2}}$ pair states. Additionally, $\ket{S_{1/2}, D_{J}}$ and $\ket{P_{J_1}, F_{J_2}}$ pair state maps are shown as they can also be coupled via 2- and 3-photon transitions from the ground state, respectively.

Pair states with strong angular dependencies between $\theta = 0, \pi/2$ can be identified by looking for bright/dark bands that shift position between the two angles (e.g. $\ket{n_1 P_{1/2}, n_2P_{1/2}}$), or dark bands that are present in just one of the two plots (see e.g. $\ket{n_1 F_{5/2}, n_2F_{7/2}}$). Additionally, a change in sign between the two angles also indicates a strong angular dependency plus an additional zero crossing of $C_6(\theta)$. Such pair states are present in every plot.

Furthermore, individual pair states can have much stronger $C_6$ values than the surrounding pair states. Usually, the F\"orster resonance lines carry such candidates. Some of these F\"orster resonant pairs possess strongly enhanced $C_6$ values if the energy defect is particularly small. Besides, one can find isolated resonances such as $\ket{27F_{7/2}, 69F_{7/2}}$ or $\ket{71S_{1/2}, 98D_{3/2}}$. All pair states which have a $C_6$ value that is at least by a factor of $\times 100$ stronger than the $C_6$ values in its vicinity are mentioned in the figure captions. The ratio to the $C_6$ values in the vicinity are usually lower for cesium than for rubidium.

The 2D map plots for all 2- or 3-photon addressable pair states are included below since the pair states for different angular quantum number combinations produce different sets of interesting features that are relevant for different applications. However, as different experiments will seek different properties in the pair states, we cannot compile an exhaustive list of interesting states. The precalculated angular channel datasets and the angular channel code are available on GitHub \citep{ARC} so that users can run their own search for interesting pair states, e.g. with different $m_j$ values or for states matching particular experimental requirements.

\begin{figure}[H]
\centering
\includegraphics[width=\linewidth]{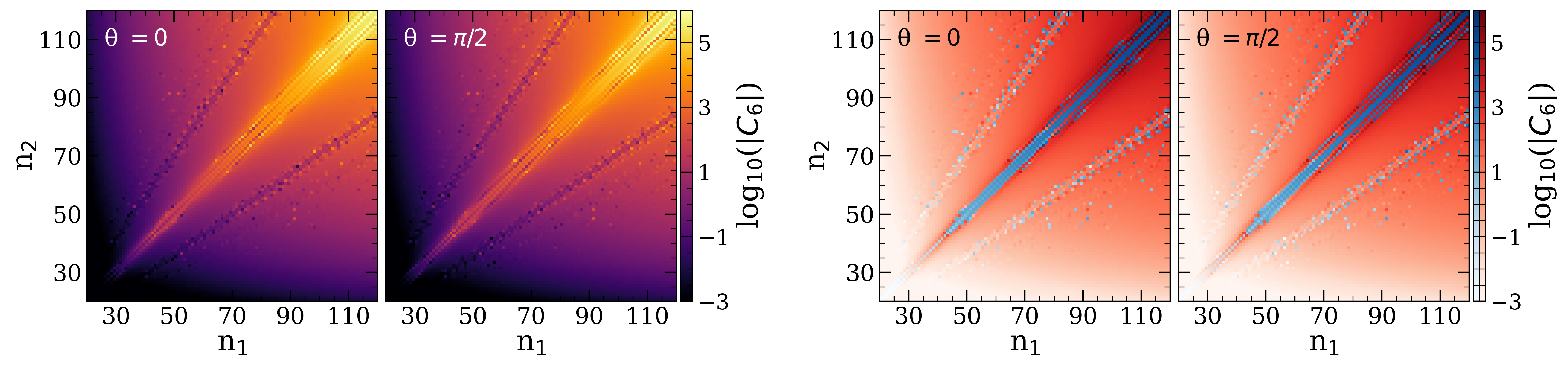}
\caption{\textbf{$\mathbf{C_6(n_1, n_2)}$ map for $\mathbf{\ket{n_1 S_{1/2},\ n_2 S_{1/2}}}$ pair states in cesium.} Absolute value (left) and sign (right) maps with $C_6 < 0$: blue, $C_6 > 0$: red. $m_1 = \pm j_1,\ m_2 = \pm j_2$ and $\tilde{m}_i = m_i$.}
\label{fig:CsS05CsS05}
\end{figure}

\begin{figure}[H]
\centering
\includegraphics[width=\linewidth]{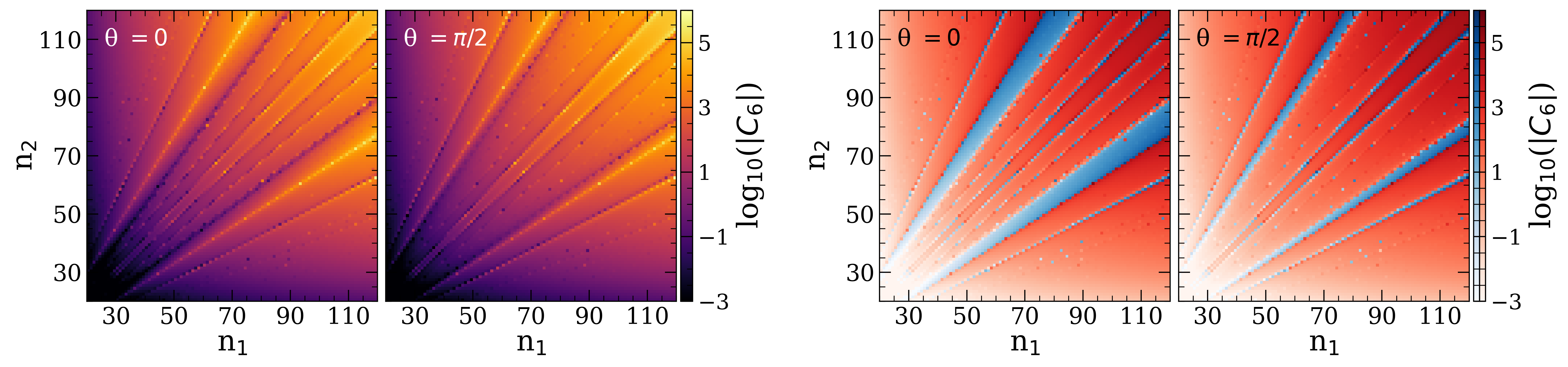}
\caption{\textbf{$\mathbf{C_6(n_1, n_2)}$ map for $\mathbf{\ket{n_1 P_{1/2},\ n_2 P_{1/2}}}$ pair states in cesium.} Absolute value (left) and sign (right) maps with $C_6 < 0$: blue, $C_6 > 0$: red. $m_1 = \pm j_1,\ m_2 = \pm j_2$ and $\tilde{m}_i = m_i$.}
\label{fig:CsP05CsP05}
\end{figure}

\begin{figure}[H]
\centering
\includegraphics[width=\linewidth]{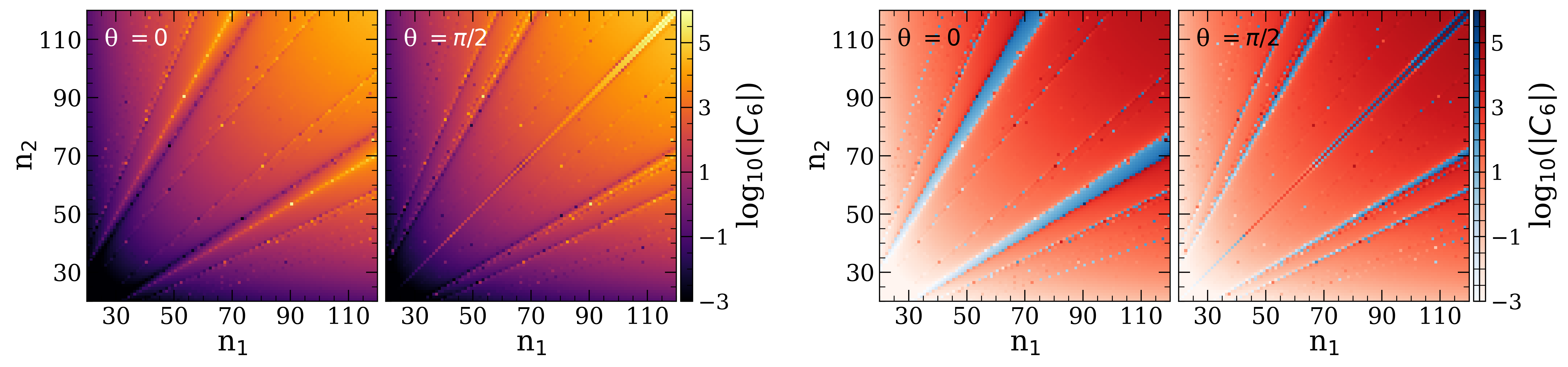}
\caption{\textbf{$\mathbf{C_6(n_1, n_2)}$ map for $\mathbf{\ket{n_1 P_{3/2},\ n_2 P_{3/2}}}$ pair states in cesium.} Absolute value (left) and sign (right) maps with $C_6 < 0$: blue, $C_6 > 0$: red. $m_1 = \pm j_1,\ m_2 = \pm j_2$ and $\tilde{m}_i = m_i$. The state (53, 90) has a $C_6$ value which is at least a factor of 831 higher than of the surrounding pair states and the pair state (40, 82) has a minimum ratio of 219.}
\label{fig:CsP15CsP15}
\end{figure}

\begin{figure}[H]
\centering
\includegraphics[width=\linewidth]{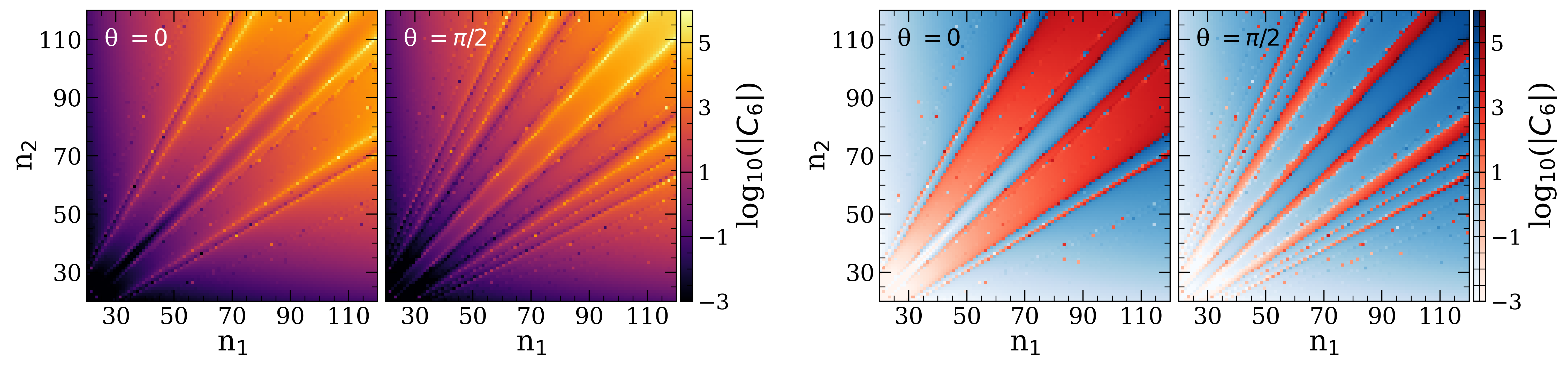}
\caption{\textbf{$\mathbf{C_6(n_1, n_2)}$ map for $\mathbf{\ket{n_1 D_{3/2},\ n_2 D_{3/2}}}$ pair states in cesium.} Absolute value (left) and sign (right) maps with $C_6 < 0$: blue, $C_6 > 0$: red. $m_1 = \pm j_1,\ m_2 = \pm j_2$ and $\tilde{m}_i = m_i$. The state (42, 71) has a $C_6$ value which is at least a factor of 576 higher than of the surrounding pair states.}
\label{fig:CsD15CsD15}
\end{figure}

\begin{figure}[H]
\centering
\includegraphics[width=\linewidth]{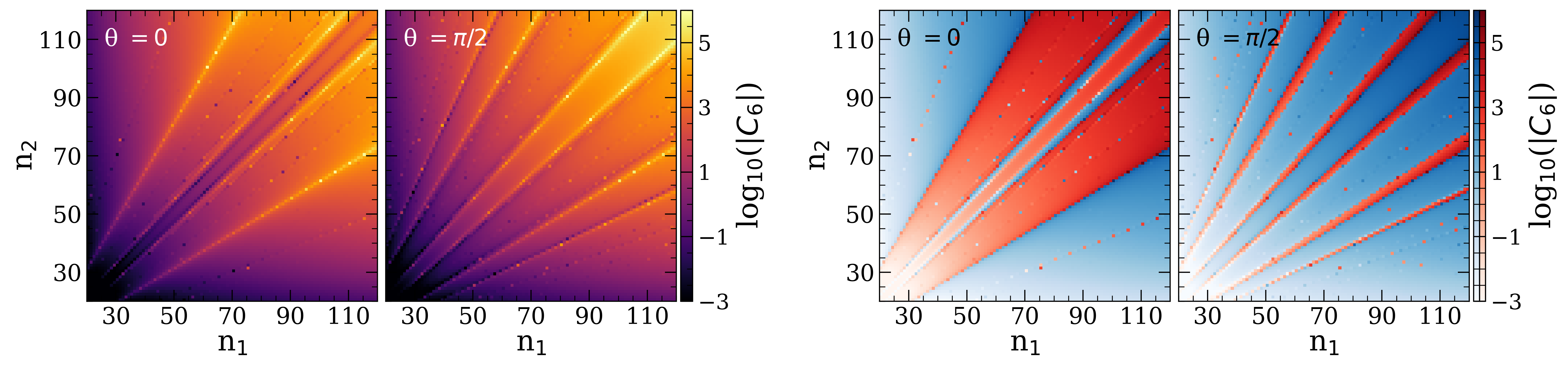}
\caption{\textbf{$\mathbf{C_6(n_1, n_2)}$ map for $\mathbf{\ket{n_1 D_{5/2},\ n_2 D_{5/2}}}$ pair states in cesium.} Absolute value (left) and sign (right) maps with $C_6 < 0$: blue, $C_6 > 0$: red. $m_1 = \pm j_1,\ m_2 = \pm j_2$ and $\tilde{m}_i = m_i$.}
\label{fig:CsD25CsD25}
\end{figure}

\begin{figure}[H]
\centering
\includegraphics[width=\linewidth]{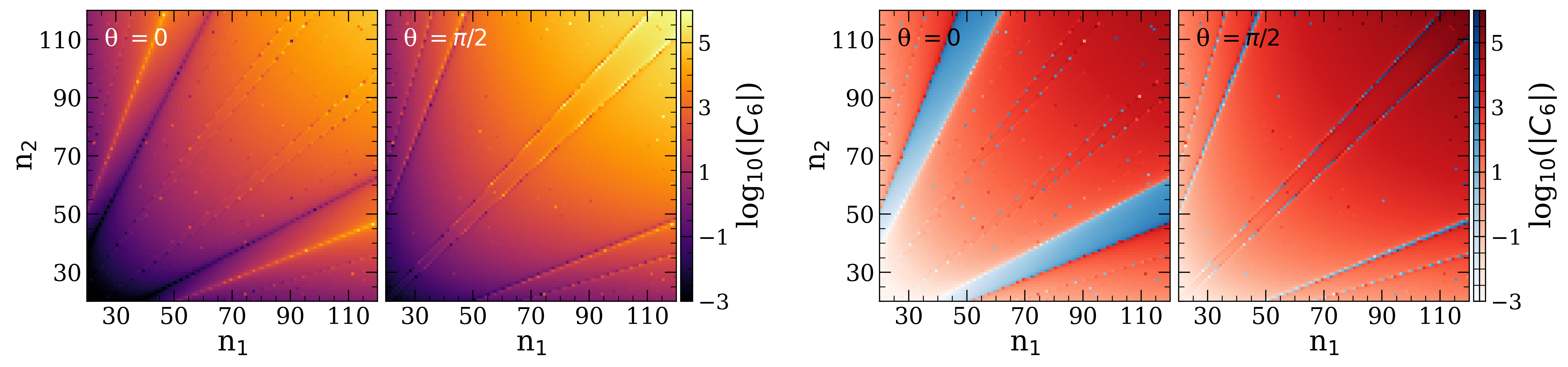}
\caption{\textbf{$\mathbf{C_6(n_1, n_2)}$ map for $\mathbf{\ket{n_1 F_{5/2},\ n_2 F_{5/2}}}$ pair states in cesium.} Absolute value (left) and sign (right) maps with $C_6 < 0$: blue, $C_6 > 0$: red. $m_1 = \pm j_1,\ m_2 = \pm j_2$ and $\tilde{m}_i = m_i$}.
\label{fig:CsF25CsF25}
\end{figure}

\begin{figure}[H]
\centering
\includegraphics[width=\linewidth]{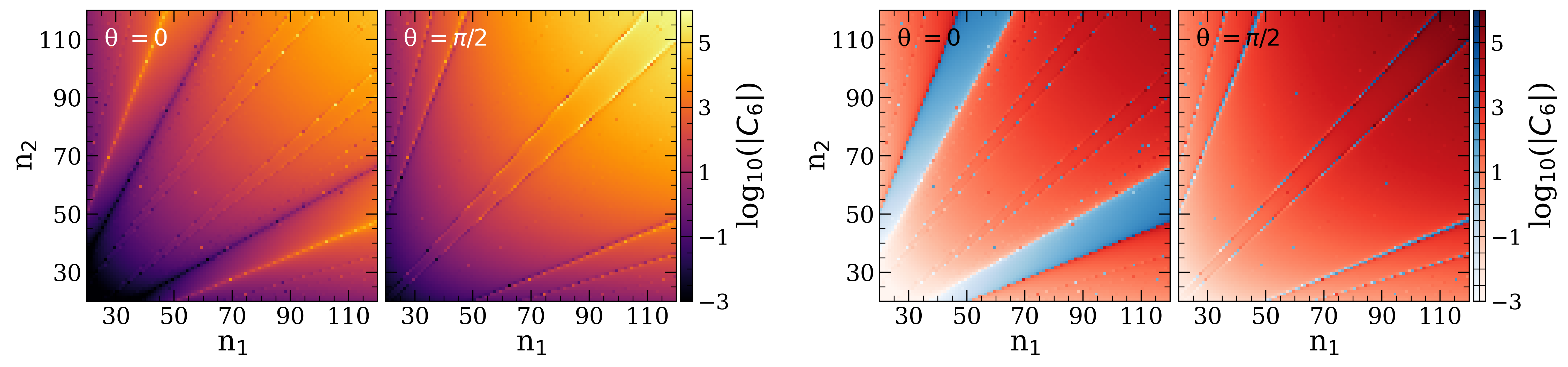}
\caption{\textbf{$\mathbf{C_6(n_1, n_2)}$ map for $\mathbf{\ket{n_1 F_{7/2},\ n_2 F_{7/2}}}$ pair states in cesium.} Absolute value (left) and sign (right) maps with $C_6 < 0$: blue, $C_6 > 0$: red. $m_1 = \pm j_1,\ m_2 = \pm j_2$ and $\tilde{m}_i = m_i$. The state (27, 69) has a $C_6$ value which is at least a factor of 178 higher than of the surrounding pair states.}
\label{fig:CsF35CsF35}
\end{figure}

\begin{figure}[H]
\centering
\includegraphics[width=\linewidth]{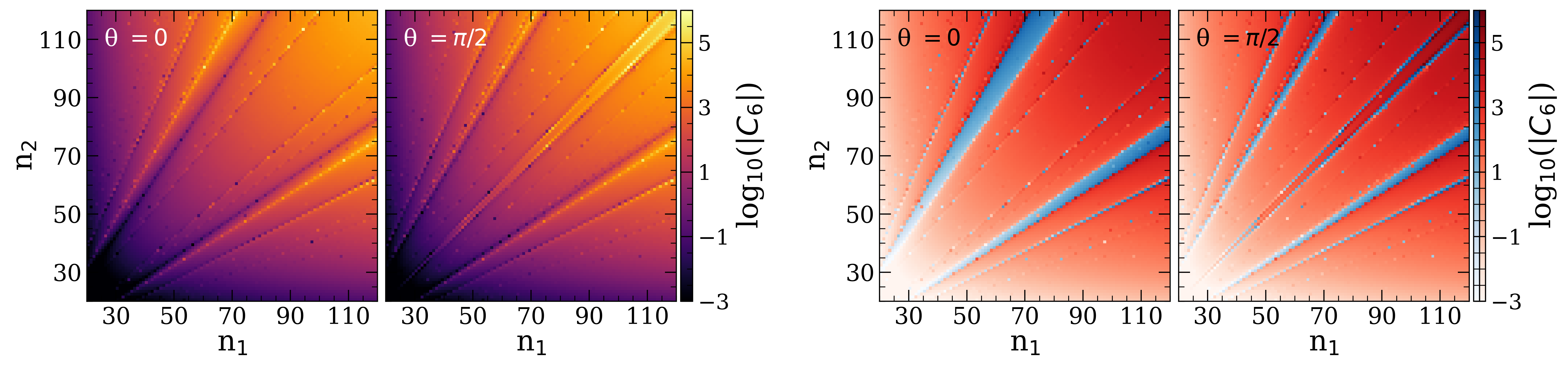}
\caption{\textbf{$\mathbf{C_6(n_1, n_2)}$ map for $\mathbf{\ket{n_1 P_{1/2},\ n_2 P_{3/2}}}$ pair states in cesium.} Absolute value (left) and sign (right) maps with $C_6 < 0$: blue, $C_6 > 0$: red. $m_1 = \pm j_1,\ m_2 = \pm j_2$ and $\tilde{m}_i = m_i$. The state (113, 94) has a $C_6$ value which is at least a factor of 102 higher than of the surrounding pair states.}
\label{fig:CsP05CsP15}
\end{figure}

\begin{figure}[H]
\centering
\includegraphics[width=\linewidth]{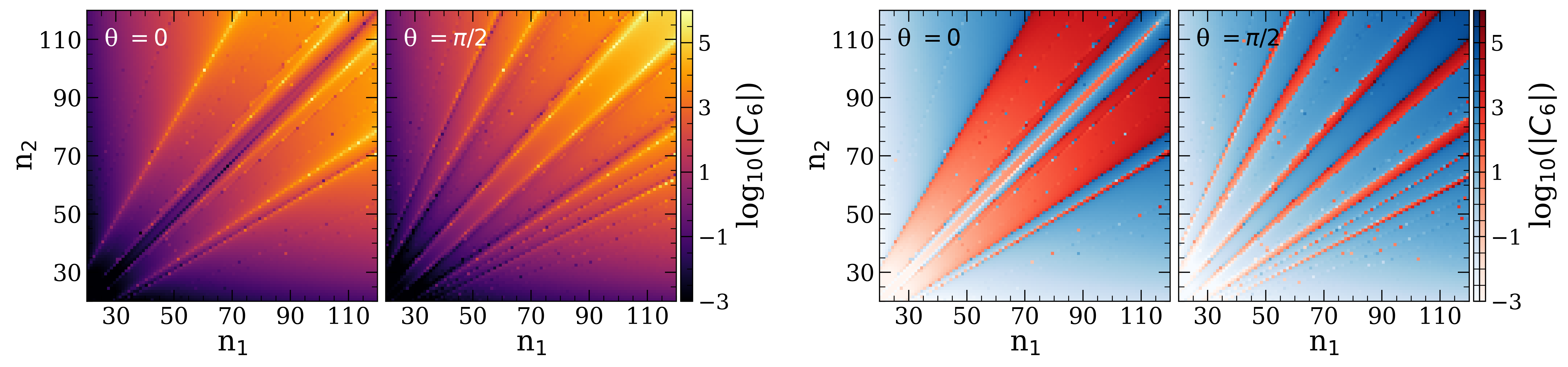}
\caption{\textbf{$\mathbf{C_6(n_1, n_2)}$ map for $\mathbf{\ket{n_1 D_{3/2},\ n_2 D_{5/2}}}$ pair states in cesium.} Absolute value (left) and sign (right) maps with $C_6 < 0$: blue, $C_6 > 0$: red. $m_1 = \pm j_1,\ m_2 = \pm j_2$ and $\tilde{m}_i = m_i$. The state (99, 60) has a $C_6$ value which is at least a factor of 167 higher than of the surrounding pair states.}
\label{fig:CsD15CsD25}
\end{figure}

\begin{figure}[H]
\centering
\includegraphics[width=\linewidth]{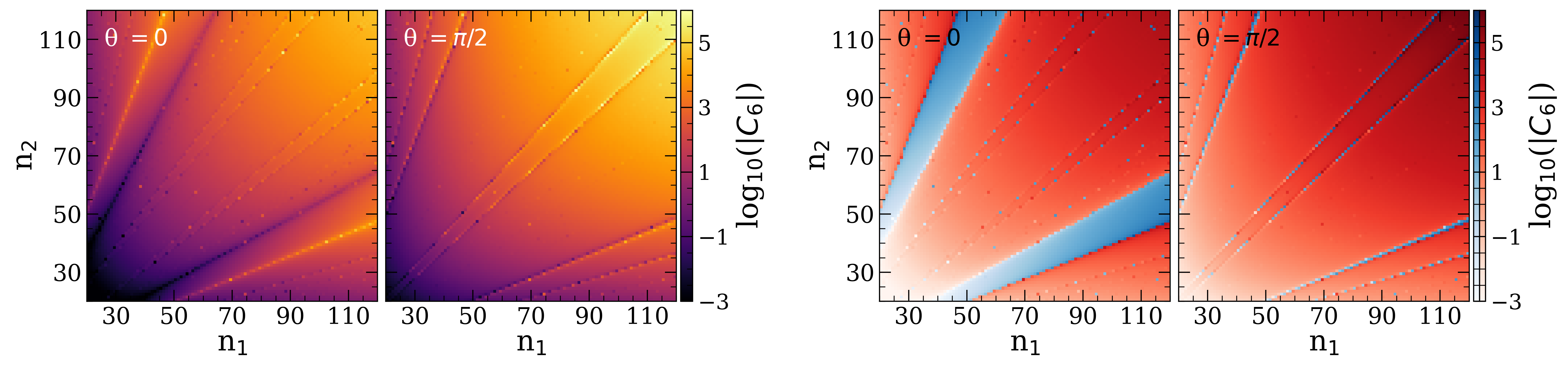}
\caption{\textbf{$\mathbf{C_6(n_1, n_2)}$ map for $\mathbf{\ket{n_1 F_{5/2},\ n_2 F_{7/2}}}$ pair states in cesium.} Absolute value (left) and sign (right) maps with $C_6 < 0$: blue, $C_6 > 0$: red. $m_1 = \pm j_1,\ m_2 = \pm j_2$ and $\tilde{m}_i = m_i$. The state (27, 69) has a $C_6$ value which is at least a factor of 181 higher than of the surrounding pair states.}
\label{fig:CsF25CsF35}
\end{figure}

\begin{figure}[H]
\centering
\includegraphics[width=\linewidth]{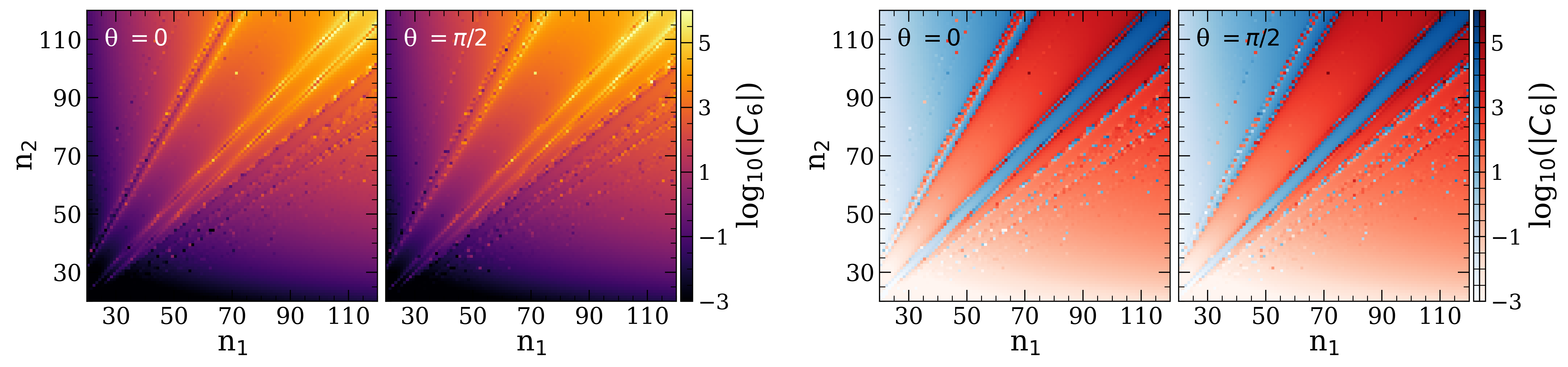}
\caption{\textbf{$\mathbf{C_6(n_1, n_2)}$ map for $\mathbf{\ket{n_1 S_{1/2},\ n_2 D_{3/2}}}$ pair states in cesium.} Absolute value (left) and sign (right) maps with $C_6 < 0$: blue, $C_6 > 0$: red. $m_1 = \pm j_1,\ m_2 = \pm j_2$ and $\tilde{m}_i = m_i$. The state (98, 71) has a $C_6$ value which is at least a factor of 246 higher than of the surrounding pair states.}
\label{fig:CsS05CsD15}
\end{figure}

\begin{figure}[H]
\centering
\includegraphics[width=\linewidth]{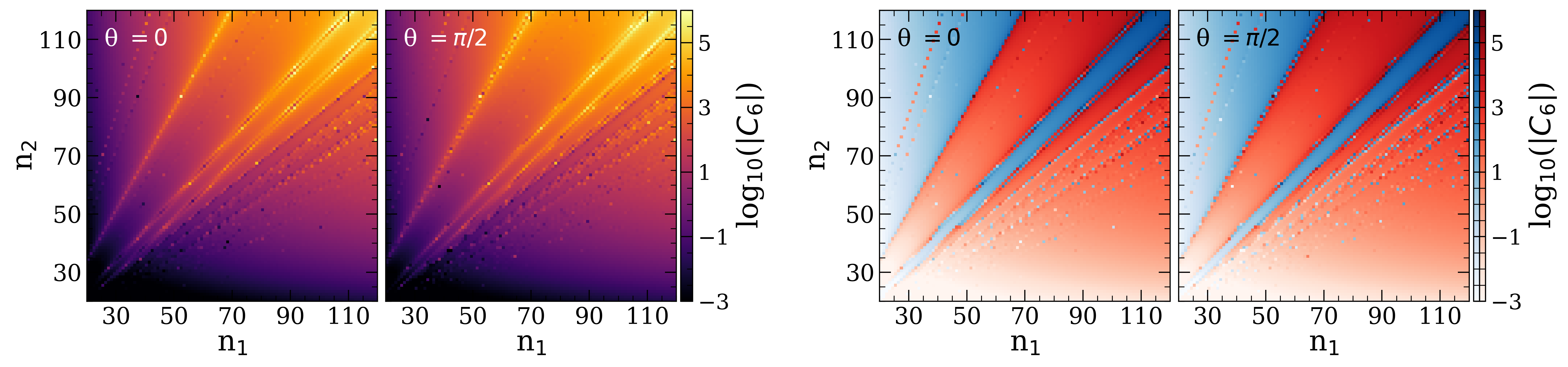}
\caption{\textbf{$\mathbf{C_6(n_1, n_2)}$ map for $\mathbf{\ket{n_1 S_{1/2},\ n_2 D_{5/2}}}$ pair states in cesium.} Absolute value (left) and sign (right) maps with $C_6 < 0$: blue, $C_6 > 0$: red. $m_1 = \pm j_1,\ m_2 = \pm j_2$ and $\tilde{m}_i = m_i$. The state (67, 78) has a $C_6$ value which is at least a factor of 187 higher than of the surrounding pair states and the pair state (90, 52) has a minimum ratio of 171.}
\label{fig:CsS05CsD25}
\end{figure}

\begin{figure}[H]
\centering
\includegraphics[width=\linewidth]{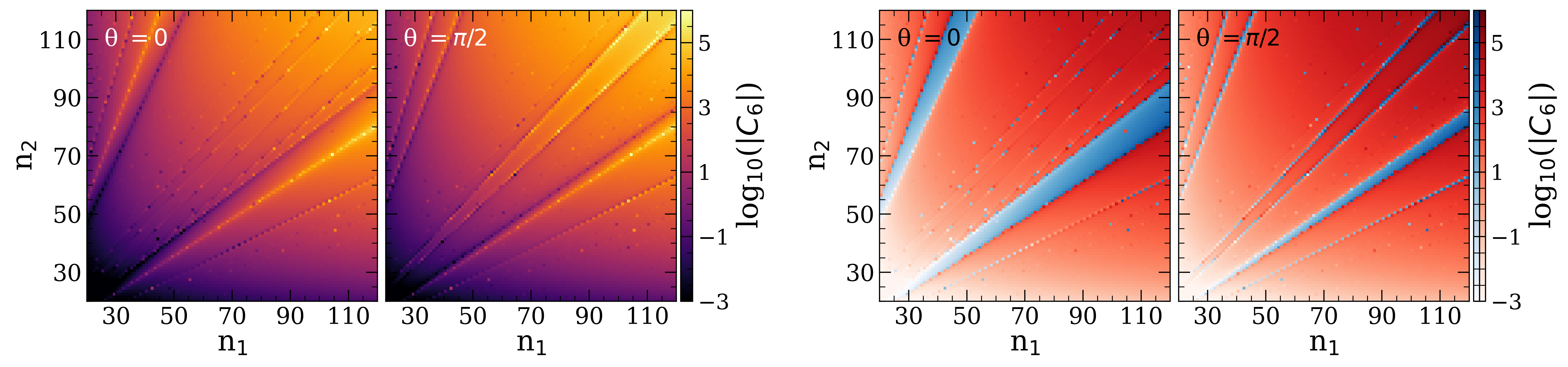}
\caption{\textbf{$\mathbf{C_6(n_1, n_2)}$ map for $\mathbf{\ket{n_1 P_{1/2},\ n_2 F_{5/2}}}$ pair states in cesium.} Absolute value (left) and sign (right) maps with $C_6 < 0$: blue, $C_6 > 0$: red. $m_1 = \pm j_1,\ m_2 = \pm j_2$ and $\tilde{m}_i = m_i$. The state (24, 42) has a $C_6$ value which is at least a factor of 629 higher than of the surrounding pair states and the pair state (54, 103) has a minimum ratio of 156.}
\label{fig:CsP05CsF25}
\end{figure}

\begin{figure}[H]
\centering
\includegraphics[width=\linewidth]{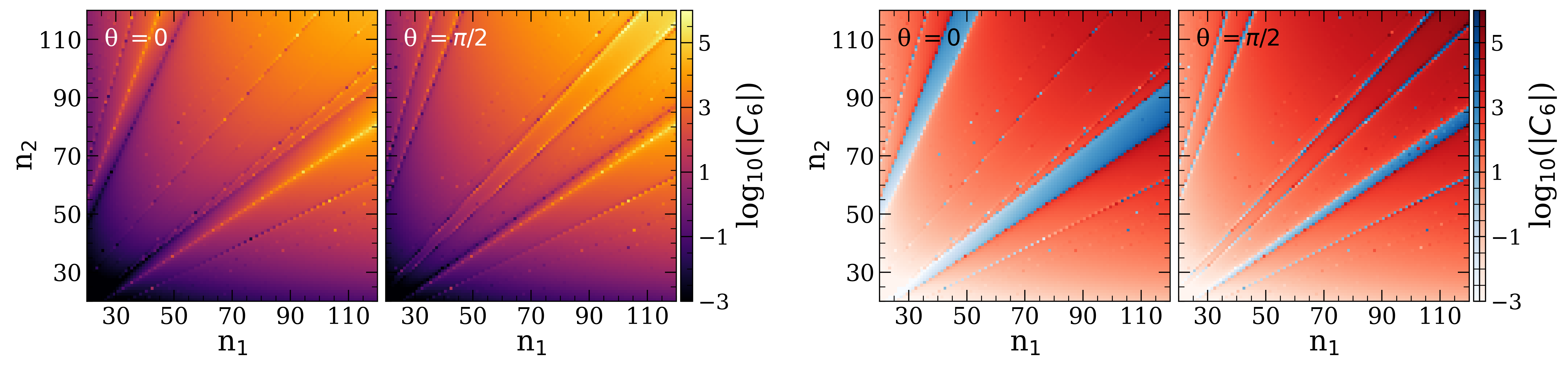}
\caption{\textbf{$\mathbf{C_6(n_1, n_2)}$ map for $\mathbf{\ket{n_1 P_{1/2},\ n_2 F_{7/2}}}$ pair states in cesium.} Absolute value (left) and sign (right) maps with $C_6 < 0$: blue, $C_6 > 0$: red. $m_1 = \pm j_1,\ m_2 = \pm j_2$ and $\tilde{m}_i = m_i$. The state (24, 42) has a $C_6$ value which is at least a factor of 662 higher than of the surrounding pair states and the pair state (54, 103) has a minimum ratio of 162.}
\label{fig:CsP05CsF35}
\end{figure}

\begin{figure}[H]
\centering
\includegraphics[width=\linewidth]{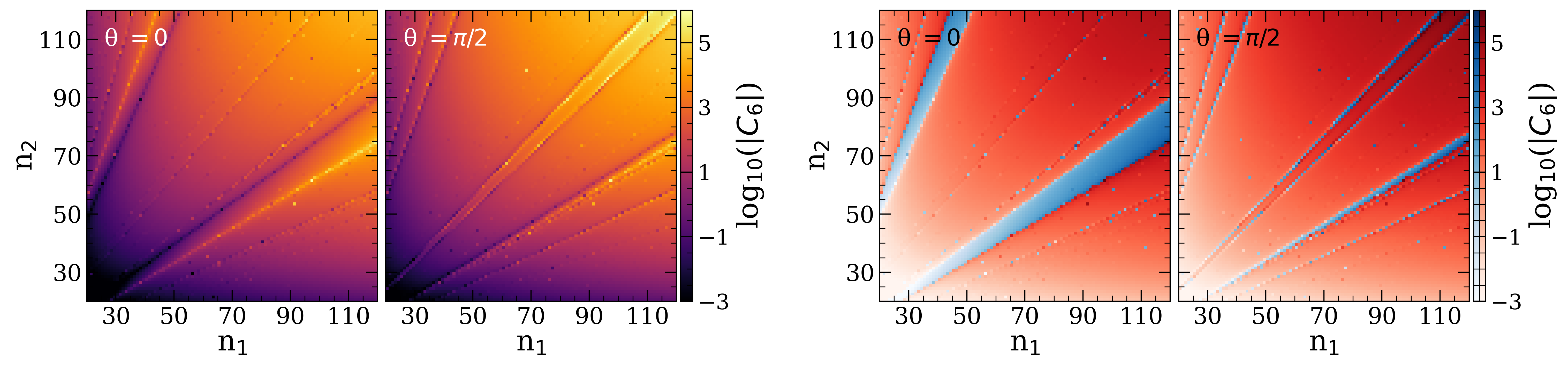}
\caption{\textbf{$\mathbf{C_6(n_1, n_2)}$ map for $\mathbf{\ket{n_1 P_{3/2},\ n_2 F_{5/2}}}$ pair states in cesium.} Absolute value (left) and sign (right) maps with $C_6 < 0$: blue, $C_6 > 0$: red. $m_1 = \pm j_1,\ m_2 = \pm j_2$ and $\tilde{m}_i = m_i$.}
\label{fig:CsP15CsF25}
\end{figure}

\begin{figure}[H]
\centering
\includegraphics[width=\linewidth]{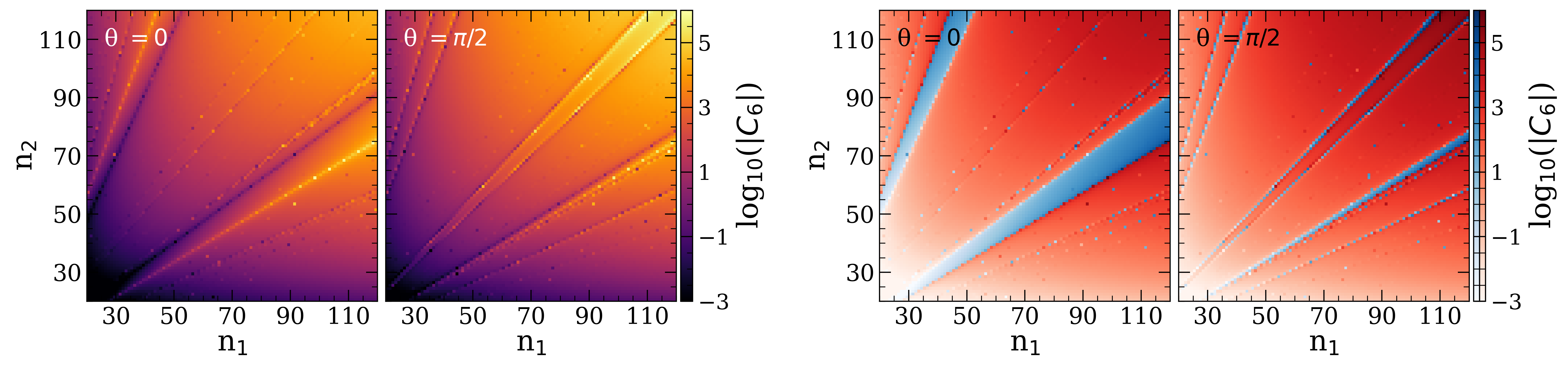}
\caption{\textbf{$\mathbf{C_6(n_1, n_2)}$ map for $\mathbf{\ket{n_1 P_{3/2},\ n_2 F_{7/2}}}$ pair states in cesium.} Absolute value (left) and sign (right) maps with $C_6 < 0$: blue, $C_6 > 0$: red. $m_1 = \pm j_1,\ m_2 = \pm j_2$ and $\tilde{m}_i = m_i$.}
\label{fig:CsP15CsF35}
\end{figure}

Note that \cref{fig:CsS05CsS05,fig:CsP05CsP05,fig:CsP15CsP15,fig:CsD15CsD15,fig:CsD25CsD25,fig:CsF25CsF25,fig:CsF35CsF35,fig:CsP05CsP15,fig:CsD15CsD25,fig:CsF25CsF35,fig:CsS05CsD15,fig:CsS05CsD25,fig:CsP05CsF25,fig:CsP05CsF35,fig:CsP15CsF25,fig:CsP15CsF35} show states with $m_i = \tilde{m_i}$ and $m_i =  j_i$, i.e. the states often chosen in experiments. These 2D maps look different for different $m_j$ states.


\twocolumngrid
\bibliography{bibliography.bib}

\end{document}